\documentclass[fleqn, preprint, 12pt, nopreprintline, authoryear]{elsarticle}

\usepackage[left=2cm,right=2cm,top=2cm,bottom=2cm]{geometry}
\usepackage[colorlinks=true,urlcolor=blue,linkcolor=blue,citecolor=blue,breaklinks=true]{hyperref}
\usepackage{natbib}
\usepackage[title]{appendix}%
\usepackage[dvipsnames]{xcolor}
\usepackage{amsmath}
\usepackage{amssymb}
\usepackage{xfrac}
\usepackage{float}
\usepackage{booktabs}
\usepackage{threeparttable}
\usepackage{colortbl}
\usepackage{hyperref}
\usepackage{adjustbox}
\usepackage{multirow}
\usepackage{multicol}
\usepackage{makecell}
\usepackage{hhline}
\usepackage{cellspace}
\usepackage[numbib]{tocbibind}
\usepackage{bm}
\usepackage{mathtools}
\usepackage{enumitem}
\usepackage{pdflscape}
\usepackage{tabularx}
\usepackage{caption}

\begin{document}
\begin{frontmatter}
\title{Interbank Decisions and Margins of Stability: an Agent-Based Stock-Flow Consistent Approach}

\author[inst1]{Jessica Reale}\corref{cor1}%

\cortext[cor1]{Corresponding author. \textit{Email address}: \href{jessica.reale@ruhr-uni-bochum.de}{jessica.reale@ruhr-uni-bochum.de}}
\address[inst1]{Chair of Macroeconomics, Faculty of Economics and Management, Ruhr-Universit\"{a}t Bochum, Universit\"{a}tsstra{\ss}e 150, 44801 Bochum (Germany).}

\date{}

\thispagestyle{empty}

\def\leftquote{`}\def\rightquote{'}
\catcode`\'=13
\def'{\ifvmode\leftquote \else \ifdim\lastskip=0pt \rightquote \else \leftquote\fi\fi}

\begin{abstract}
\sloppy

This study investigates the functioning of modern payment systems through the lens of banks‘ maturity mismatch practices, and it examines the effects of banks‘ refusal to roll over short-term interbank liabilities on financial stability. Within an agent-based stock-flow consistent framework, banks can engage in two segments of the interbank market that differ in maturity, overnight and term. We compare two interbank matching scenarios to assess how bank-specific maturity targets, dependent on the dictates of the Net Stable Funding Ratio, impact the dynamics of the interbank market and the effectiveness of conventional monetary policies. The findings reveal that maturity misalignment between deficit and surplus banks compromises the interbank market's efficiency and increases reliance on the central bank's standing facilities. Monetary policy interest-rate steering practices also become less effective. The study also uncovers a dual stability-based configuration in the banking sector, resembling the segmented European interbank structure.  This paper suggests that heterogeneous maturity mismatches between surplus and deficit banks may result in asymmetric funding frictions that might precede credit- and sovereign-risk explanations of interbank tensions. Also,
a combined examination of macroprudential tools and rollover-based interbank dynamics can enhance our understanding of how regulatory changes impact the stability of heterogeneous banking sectors.

\end{abstract}

\end{frontmatter}
\sloppy
\section{Introduction}
\sloppy
Interbank lending practices have come to the forefront of financial stability debates after the global financial crisis (GFC). Banks' inability and/or unwillingness to exchange liquidity in the interbank market – to smooth liquidity shocks and to clear interbank positions – exposed the daily functioning of modern payment systems to "the emergence of contagion and systemic risk" \citep[][p. 5]{iori2008network}. Before 2008, banks' assets and liabilities were analyzed individually, disregarding any potential contagion effect. After the crisis, the need to abandon the usual microprudential focus and to analyze the formation of endogenous interbank networks became crucial because of (i) highly interconnected banking institutions and (ii) interdependent interbank liability exposures. As a result, the notion of systemic risk became widely accepted as being "an intrinsic property of modern economies" \citep[][p. 3]{lengnick2013agent}, where the interbank market essentially plays the role of \textit{economy stabilizer} in normal times and of \textit{shocks amplifier} during crises. The GFC has been characterized by a condition of \textit{systemic illiquidity} that arose from the "inherent liquidity mismatch" \citep[][p. 2]{ferrara2019systemic} in banks' balance sheets, which led to a cascade of funding shortfalls. Banks' inability to rollover their short-term interbank exposures thus had a greater impact than  banks' defaults in the propagation of shocks and contagion risk \citep{ferrara2019systemic}. Indeed, systemic instability has been mainly triggered by excessive maturity mismatch practices, which increased the risk of rollover borne by banks and gave an endogenous boost to the GFC \citep{lengnick2013agent}.
The inability of banks to re-access the market to pay for existing debts – i.e., to  rollover their short-term exposures – amplified the initial liquidity shock and led to the emergence of a \textit{fragile} financial system \citep{montagna2016multi}.\footnote{The cyclical tendency of the capitalist system to alternate from a state of robust finance to a state of fragile finance, theorized by Minsky in his \textit{Financial Instability Hypothesis} \citep{Minsky.1994}, can be interpreted as driven by banks' refinance mechanisms and their debt maturity structure decisions \citep{Reale.2023}.} As a result, the vulnerabilities behind banks' refinance mechanisms on the interbank market became crucial to analyzing the potential spread of contagion and systemic risks in modern capitalist systems.

Higher rollover risks increased borrowing costs in the term segment of the unsecured money market and triggered the interbank market freeze characterizing the financial crisis \citep{acharya2011model}. Indeed, the volumes exchanged on the term segment of the European interbank market have substantially decreased after the collapse of Lehman Brothers \citep{abbassi2013network}. This event suggests that "focusing on a single interbank segment can underestimate shocks propagation" \citep[][p. 5]{Bargigli.2015}  when analyzing the impact of the interbank lending network on financial stability. 
The GFC thus revealed that (i) contagion and systemic risks result from banks' excessive maturity mismatch practices, (ii) agents' exposure to frequent refinancing needs are consolidated features of the functioning of the banking system, and (iii) the conduct of the overnight and the term interbank segments cannot be analyzed individually.

In light of these considerations, this paper aims to (i) examine the efficiency of the interbank market when banks can choose the maturity structure of their funding sources and limit their exposures to excessive maturity mismatches and the risk of rollover and to (ii) analyze how the propagation of maturity shocks among banks affects their clearing operations within the theorized payment system. We thus use our model to answer the following research questions: What are the implications of limiting the exposures to excessive maturity mismatches and the risk of rollover on the clearing operations of the interbank market?
Under which conditions, if any, do mismatched maturity preferences between surplus and deficit banks impair the functioning of the interbank market and central banks' policies?
To pursue these objectives, this article proposes an Agent-Based Stock-Flow Consistent (AB-SFC) analysis of a complex evolving system in line with recent modeling practices \citep[see][among others]{Caiani.2018, Grabner.2022, Schutz.2022}. On the one hand, Agent-Based Models (ABM) are best suited to model such a framework, given the high complexity and heterogeneity of money markets \citep{ Wolski.2016, Berndt.2019}.  On the other hand, Stock-Flow Consistent (SFC) applications à la \cite{Godley.2006} allow for the interaction between the real and financial sides of the economy, whose institutions are explicitly modeled \citep{Ioannou.2018}.  By combining these two methods and modeling a heterogeneous banking sector, this model mimics the functioning of Target 2 (T2) and captures the interaction of the various maturity-based interbank segments. In this paper, two groups of banks interact with each other \citep{Reale.2022}. While one group grants credit only to households – i.e., commercial banks – business banks provide loans to firms to finance production. Therefore, for every consumption and/or wage payment decision, banks may suffer from a payment outflow that must be settled in the central bank's money. To do so, banks can access a multilayered unsecured interbank network diversified by maturity, which is crucial to assess financial stability. Indeed, the interbank network is composed of two segments, the overnight and the term one, to model the behavior of banks when they have  "preferences for diversified funding sources in terms of rollover risk" \citep[][p. 2]{halaj.2015}. We model two interbank matching mechanisms to isolate the impact of rollover risk on interbank dynamics. The "Baseline" matching protocol features deficit banks looking for partners that can fully accommodate interbank requests in volumes and maturity. Instead, the "Maturity" scenario introduces 
endogenous target financial ratios deriving from the guidelines imposed by the \textit{Net Stable Funding Ratio} (NSFR) of Basel III, such that interbank matching depends on the residual maturities of assets and liabilities in the balance sheets of surplus and deficit banks.

The outline of this article is as follows. Section \ref{section 2} presents the literature related to this study. Section \ref{section 3} introduces the general characteristics of the model along with the mathematical formalization of banks' behaviors and interbank market matching mechanisms. The next section (section \ref{section 4}) describes the simulation strategy and discusses the experiments and the main results. The last section concludes.

\section{Related literature}\label{section 2}
This paper is closely aligned with two distinct bodies of literature. Firstly, we summarize contributions investigating the interbank market through network analyses or multi-agent perspectives. Within this body of literature, specific attention is given to articles that examine the different segments of the interbank network, taking into account maturity diversification and analyzing systemic risk by considering banks' funding liquidity risk and excessive maturity mismatch. Secondly, we relate to studies that employ the AB-SFC methodology to examine the dynamics of the banking sector.

Among the first group of contributions, \cite{halaj.2015} provide a combination of agent-based modeling and theoretical sequential games to study how interbank networks endogenously emerge and impact contagion risk. The authors adopt a \textit{portfolio optimization} perspective by distinguishing between an asset-side optimization – based on regulatory constraints – and a liability-side optimization, dependent on funding liquidity risk, such that banks choose their preferred funding structures to limit their exposure to the risk of rollover.\footnote{In this context, rollover risk depends on lending banks' probability of default.  When a creditor goes bankrupt, the inability to roll over materializes in the form of funding risk, measured as the variance of the portfolio.} Despite funding liquidity risk arising from short-term funding of long-term lending practices (maturity mismatches), the authors do not explicitly model the maturity structure of interbank assets and liabilities, i.e., they assume that all the items in banks' balance sheets have the same residual maturity.
\cite{Bargigli.2015} analyze maturity-varying credit relations by focusing on a series of interbank layers that provide a more accurate representation of the interbank market. Their model studies the Italian interbank market by adopting a \textit{multiplex} or \textit{multi-layer} approach, which exploits the supervisory reports of Banca D'Italia.

This methodology becomes necessary to avoid focusing only on the overnight unsecured interbank market segment. Along these lines, \cite{montagna2016multi} adopt a multi-layered interbank network analysis to investigate systemic risk in the highly interconnected European financial system. In their model, banks can interact in different layers diversified by maturity, with short-term loans reflecting funding risk and long-term loans counterparty credit risk. \cite{popoyan2017taming} build an ABM with interbank market dynamics to investigate the interplay between monetary policy tools and macroprudential regulation measures. They analyze 
the emergence of endogenous banking crises and interbank frictions by focusing on the \textit{Liquidity Coverage Ratio} (LCR) of Basel III, mixed with several policy corridor regimes. In this model, banks' demand for funds on the interbank market originates from prudential regulation purposes, i.e., banks are either interbank lenders or borrowers according to whether they meet or not the LCR.\footnote{When the LCR is not satisfied, banks are in shortage of liquidity; the opposite happens when the regulatory constraint is satisfied.}
Another multi-agent model that studies how banks' decisions affect the structure of the interbank market has been provided by \cite{Liu.2018}. In their model, the authors distinguish between three types of interbank contracts diversified by maturity (overnight, short-term, and long-term) and between two types of banks diversified by size (small VS large ones). The formation of new debts – or links – in the interbank network depends on the maturity of the contract.
In their model, potential borrowing banks demand funds for a specific maturity based on financial targets: if the overnight (or short-term or long-term) borrowing ratio is lower than the target one, they will search for overnight (or short-term or long-term) funds on the interbank market. Lending banks' decisions about whether or not to accommodate the demanded duration will depend on lending ratios for maturities and on a \textit{score mechanism} that captures banks' tendency to keep existing relationships. Last but not least, \cite{ferrara2019systemic} study systemic illiquidity arising from rollover failures in the interbank network and investigate the effectiveness of macroprudential measures in minimizing this issue. The authors focus on the potential inability of banking institutions to repay their obligations when they become due and reveal that the propagation of a shock becomes stronger when the interbank funding network suffers from a cascade of funding shortfalls rather than a cascade of banks' defaults.

This paper is also linked to the contributions that adopt AB-SFC models.
 \cite{lengnick2013agent} build a simple AB-SFC model of a monetary economy where the banking sector is seen as a large decentralized economic system. In their attempt to study the effects of interbank lending on financial stability, the authors reveal the existence of potential instability that follows from the maturity mismatch of assets and liabilities in banks' balance sheets. However, they model the endogenous creation of money as deriving from "behavioral interactions" \cite[][p. 16]{lengnick2013agent}, in contradiction with the post-Keynesian concept of endogenous money adopted in the present analysis.\footnote{Money is endogenously created by the private sector out of thin air. As a result, the causal link runs from loans to deposits to reserves, opposite to what is postulated by standard monetary theories.}
Potential financial frictions are also analyzed by \cite{caiani2016economics} in their \textit{benchmark model for macroeconomics}. In their AB-SFC framework, the authors distinguish between two types of firms, capital and consumption ones, and introduce a mechanism for banks to actively manage their balance sheets to supply credit to firms and households. This mechanism allows (i) modeling \textit{endogenous evolving strategies} according to which banks manage the interest rates charged on the credit market and (ii) introducing a \textit{case-by-case} quantity rationing procedure that accounts for the risk and the expected internal rate of return of each credit application. \cite{schasfoort2017monetary} extend this framework to interbank market dynamics. Their model attempts to test the strength of the various channels of the monetary policy transmission mechanism following an increase in the central bank's policy rate. In this model, banks determine the rates of interest charged on the credit market as dependent on the funding costs they face on the interbank market. By doing so, they found out that increasing policy rates lead to an interbank shock which manifests itself in the form of increased funding costs and triggers (i) higher rates on loans to firms and households, (ii) lower rates of investments, and (iii) inflation. 
Last, \cite{reissl2018monetary} adopts a more hybrid technique by modeling a disaggregate banking sector – leaving all other sectors in aggregate terms – to study how banks form their heterogeneous expectation on the credit market. 

This paper contributes to these two bodies of literature in two significant ways. Firstly, it designs more intricate banks' decisions regarding debt maturity structure in the interbank market, surpassing the limitations of existing AB-SFC models where banks' behavior is "modeled in a fairly simple fashion" \citep[][p. 19]{reissl2018monetary}. Secondly, it explicitly incorporates the endogenous creation of money in line with the principles of post-Keynesian monetary theory, upon which the SFC framework is built.

\section{The model}\label{section 3}
\begin{figure}[t]
    \centering
\includegraphics[scale=0.8]{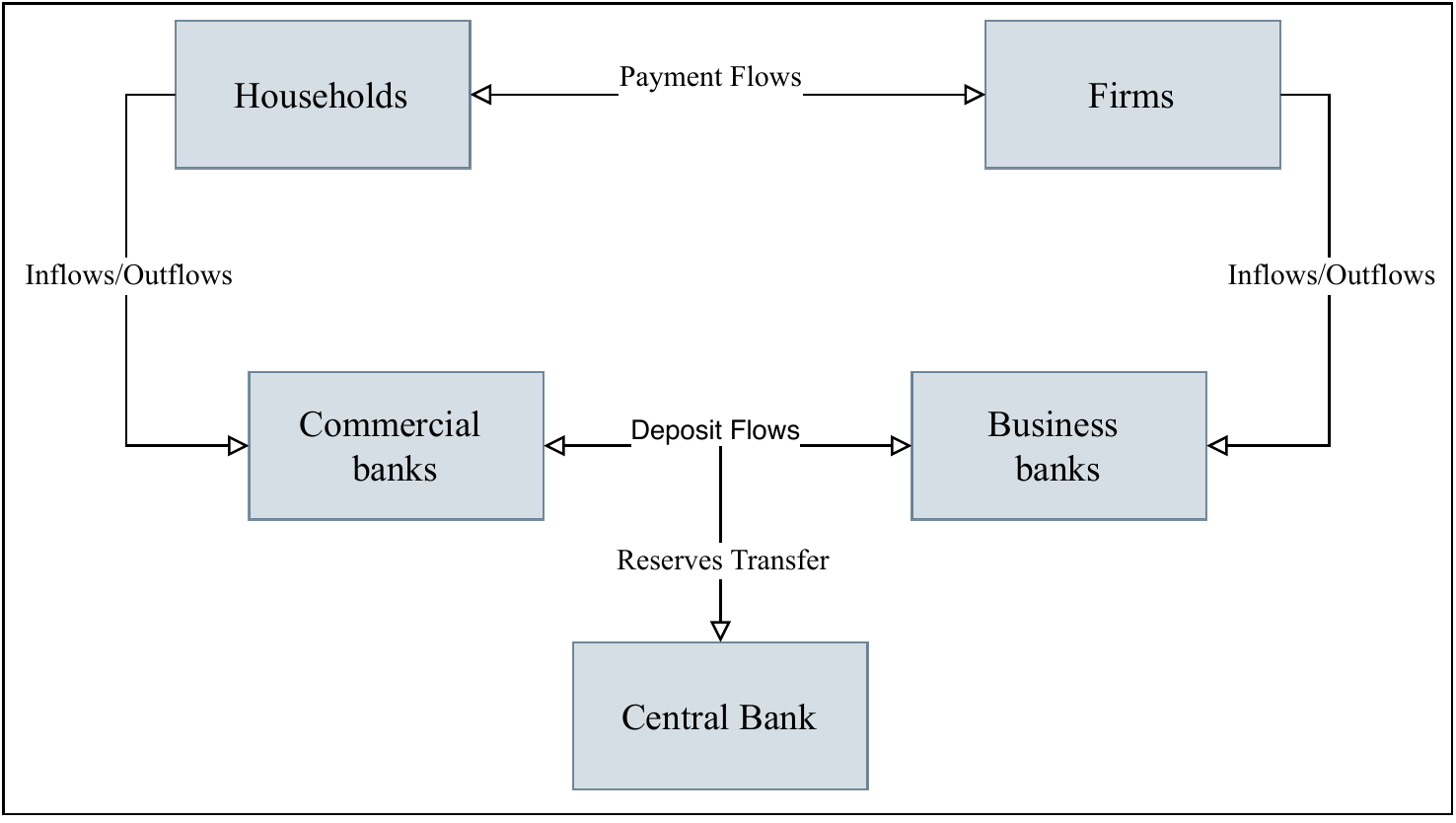}
    \caption{Monetized Production Economy.}\label{fig:figure1}
\end{figure}

\sloppy
The postulated economy comprises one government, one central bank, and a collection of households, firms, and banks.
Households ($hh=1,2,...,N_{h}$) buy consumption goods from firms, save in the form of banks' deposits ($D_{hh}$), receive wage payments and dividends, pay taxes to the government and have access to banks' loans ($L_{hh}$). 
Firms ($f=1,2,...,N_{f}$) receive loans ($L_{f}$) to finance the production of a homogeneous consumption good and "are assumed to manage their liquidity by holding bank deposits" \citep[][p. 9]{michell2014steindlian} – $D_{f}$. 
The public sector is composed of the government and the central bank. The former collects taxes paid by households and finances its public debt by issuing short-term securities ($B$) – which are bought by the banking sector ($B{b}$) and, residually, by the central bank ($B{cb}$) – and long-term bonds ($B^{lr}$) to absorb banks' non-performing loans ($NPL$). The central bank provides required reserves ($HPM$), advances ($A$), and standing facilities that banks can use when interbank exchanges come to a standstill ($R^{l}$ and $R^{d}$).
The most detailed part of this model concerns the behavior of banking institutions. The banking sector comprises two collections of banks to resemble the functioning of a potential payment system within an overdraft economy and simulate the operational framework of the European T2. For this reason, commercial banks ($j=1,2,...,N_{bj}$) grant loans only to households, and business banks $k$ ($k=1,2,...,N_{bk}$) provide funds only to firms to finance production. This fragmentation of the banking sector allows capturing the dynamics of a monetized production economy as stylized in Fig. \ref{fig:figure1}. 
Whenever households buy consumption goods or receive wage payments from firms, the payment flows between these two agents translate into deposit outflows and inflows into banks' balance sheets, whose debt relationship must be cleared through a transfer of reserves in the liability side of the central bank's balance sheet. 
When banks do not have enough reserves to cover the payment transfer arising from the initial monetary transaction, they access the unsecured interbank market. Two segments of the unsecured interbank market are considered: the overnight and the term one.  This specification allows capturing banks' debt maturity structure decisions as dependent on the level of maturity mismatch, which, if excessive, might be responsible for potential refinancing vulnerabilities in the form of a high risk of rollover.

In each simulation period, these agents interact in four markets: households interact with firms on the  consumption goods market; firms and households  interact with banking institutions on the credit market; banks and the central bank interact with the government on the market for securities – bills, and bonds; last, the two types of banks interact with each other on the interbank market.
While bills market interactions depend on simple quantitative buffers – standard in SFC models – and the government accommodates banks' demand for bonds, the interactions in the credit and goods markets take place through a common matching protocol \citep{michell2014steindlian, caiani2016economics, bargigli2016simple, schasfoort2017monetary}.
The mechanism for partners' selection in the credit and goods markets is random at $t_{0}$ \citep{gatti2010financial}. From the subsequent periods on, demand agents (households and firms) observe the interest rates and prices charged by a random subset of supply agents (banks and firms) and decide whether to choose the current supplier or to switch to a new one according to a \textit{probability of switching} partners and an exogenous \textit{intensity of choice} parameter \citep{gatti2010financial, schasfoort2017monetary}. 
Banks can exchange interbank funds in two segments diversified by maturity, overnight, and term.
We model two interbank matching protocols that allow us to distinguish between a \textit{baseline} and a \textit{maturity-based} scenario. First, we set a "Baseline" matching where banks engage in interbank exchanges with partners who can best accommodate their customers' demand for funds in both segments.  Second, we implement a "Maturity" scenario where banks engage in an \textit{active search for counterparties}  \citep{Liu.2018} based on the maturities of the exchangeable interbank contracts, see section \ref{sec: matching}.
The balance sheet matrix – table \ref{balmatrix} – depicts the stocks held by each collection of agents in each sector.

\begin{table}[t]
    \centering
   \resizebox{!}{.16\paperwidth}{\begin{tabular}{|llllll|l|}
    \hline
    & Firms & Households & Banks & Government & Central bank & $\Sigma$\\
    \hline \hline
        Capital & $+K$ & & & & & $+K$\\
        Inventories & $+INV$& & & & & $+INV$\\
        Loans & $- L_{f}$ & $-L_{h}$& $+L$& & & 0 \\
        Deposits & $+Df$& $+Dh$ & $-D$ & & & 0 \\
         Bills & & &$+Bb$ & $-B$ & $+Bcb$&  0\\
        Bonds & & &$+B^{lr}$  & $-B^{lr}$ & & 0\\ \hline \hline
        High powered money & &&$+HPM$ & &$-HPM$ &0 \\
        Advances & &&$-A$ & &$+A$ &0 \\
        Lending Facility &&&$-R^{l}$& &$+R^{l}$ &0\\
Deposit Facility &&& $+R^{d}$  && $-R^{d}$& 0\\
        \hline \hline
      \textit{Balance} & $-NWf$& $-NWh$ & $-NWb$& $+ GD$ &0& $-(K + INV)$ \\ 
      $\Sigma$ & 0 & 0 & 0  & 0 & 0& 0 \\
      \hline
    \end{tabular}}
    \caption{Balance sheet matrix.}
    \label{balmatrix}
\end{table}

\subsection{Sequence of Events}
\begin{figure}[t]
    \centering
    \includegraphics{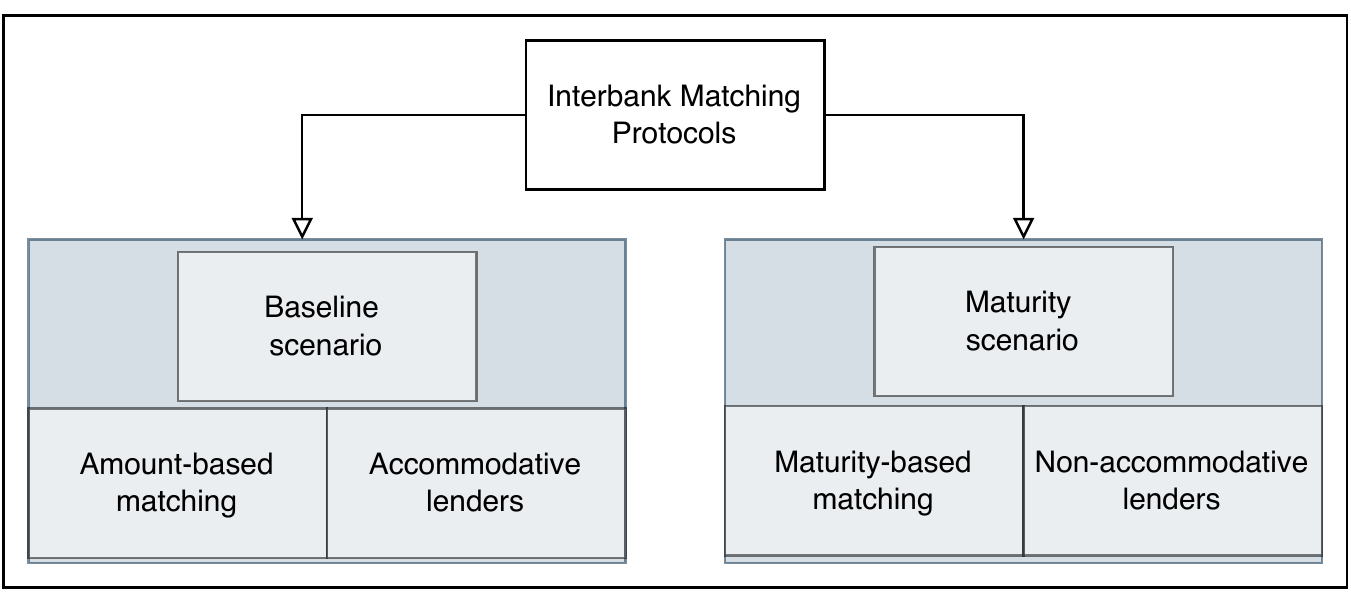}
    \caption{Interbank matching and scenarios.}
    \label{fig:figure2}
\end{figure}

At each simulation round, the following order of events takes place:

\begin{enumerate}[noitemsep]
\item \textit{Pricing:} firms set prices as a markup over past unit direct costs. Banks update the interest rates for loans and deposits according to previous-period funding costs;

\item \textit{Production planning:} firms set their desired level of investments, make production decisions based on expected sales and short-run inventory targets, pay wages, and produce output;
\item \textit{Consumption goods market:}  households observe the prices on consumption goods charged by a subset of firms and decide whether to change partner or buy from the previous matched firm. They make consumption decisions based on expected income and pay taxes on wages. If sales exceed output and inventories, consumers are rationed accordingly.

\item \textit{Credit market:} firms and households observe a random subset of banks and forward their loan applications to the bank charging the lowest interest rate. A constant proportion of loans is assumed to be non-performing each period;

\item \textit{Interbank market:} banks update their interbank status – \textit{surplus}, \textit{deficit} or \textit{neutral} – according to the payment flows arising from their customers' consumption spending and wage payments.  Banks engage in interbank transactions on the overnight and term segments using two matching protocols (see Fig. \ref{fig:figure2}). When interbank transactions freeze, banks can access the central bank's standing facility to acquire the reserves needed or deposit their excess holdings. Banking institutions also update their funding costs based on interbank transactions and their recourse to the lending facility;

\item \textit{Securities market:} the government issues long-term bonds – to absorb banks' non-performing loans – and bills, and the central bank clears the bills market.
\end{enumerate}

The following sections describe banks' behaviors only, as the behavior of the rest of the economy is standard in the SFC literature.\footnote{All the exogenous variables are denoted with a bar.}

\subsection{Banks' behavior}
Both types of banks ($i = \{j,k\}$) receive high-powered money from the central bank to meet their reserve requirements, defined as a fixed proportion ($\bar{\mu} + \bar{v}$) of current deposits  (Eq. \ref{eq:hpm}). They buy government bills as a buffer,  constrained by the amount the government issues (Eq. \ref{eq: Bb}). If bills are null, the central bank's advances (Eq. \ref{eq: adv}) act as buffer variable \citep{dafermos2012liquidity}. At each period, a constant proportion of firms' and households' loans is assumed to be non-performing. Non-performing loans ($NPL$) are transferred to the government through long-term securities (Eq. \ref{eq: Blrb}), and banks' profits (Eq. \ref{eq: Pb}) are distributed to households.
\begin{align}           
    &HPM_{i,t}=(\bar{\mu} + \bar{v})D_{i,t}\label{eq:hpm}\\
    & Bb_{i,t} = \text{min}(D_{i,t} + NPL_{i,t} + R^{l}_{i,t} - L_{i,t} - \bar{\mu}D_{i,t} - B^{lr}_{i,t} - R^{d}_{i,t}, \; \frac{B}{N_{b}})\label{eq: Bb}\\
     & A_{i,t} = \left \{\begin{array}{ll}
     L_{i,t} + R^{d}_{i,t} + HPM_{i,t} + B^{lr}_{i,t} - D_{i,t} - NPL_{i,t} - R^{l}_{i,t}
        & \text{if}\; Bb_{i,t} < 0  \\
      \bar{v} D_{i,t} \;\;\; \text{otherwise} &
    \end{array} \right . \label{eq: adv}\\
    & B^{lr}_{i,t} = NPL_{i,t} \label{eq: Blrb}\\
    &Pb_{i,t} = i^{l}_{i,t-1} L_{i,t-1} + \overline{icb}^{t} HPM_{i,t-1} + \bar{ib}Bb_{i,t-1} + \bar{ib}^{lr} B^{lr}_{i,t-1} + \overline{icb}^{d}R^{d}_{i,t-1} - i^{d}_{i,t-1} D_{i,t-1} - \label{eq: Pb} \\ \notag 
    & - \overline{icb}^{t}A_{i,t-1} - \overline{icb}^{l}R^{l}_{i,t-1} 
\end{align}

Banks access the interbank market to settle the real sector's transactions as their demand for interbank loans originates from \textit{payment settlement purposes} \citep{Reale.2022}. 
Banks' roles in the interbank market thus depend on outflows and inflows (Eq. \ref{eq: flow}) borne by banks due to the conduct of business in the real sector.
Therefore, every consumption decision and wage payment triggers a \textit{reserve adjustment process} that defines the pool of interbank lenders (surplus banks) and borrowers (deficit banks). Commercial (business) banks have a set of households' (firms') customers $S_{i,t} = \{i\; | i \in N_{h}\; \text{and}\; \text{link}_{i, x} \neq 0\}$ ($S_{i,t} = \{i\; | i \in N_{f}\; \text{and}\; \text{link}_{i, x} \neq 0\}$) that is updated every period according to the credit market matching protocol. As such, banks' payment flows are indirectly affected by the interest rates they charge on loans. 
Banks incurring a payment outflow ($f < 0$) ask for interbank funds ($DF$) if their change in high-powered money is not enough to smooth the liquidity shock (Eq. \ref{eq: ib-d}). The amount of reserves lending banks can provide on the interbank market ($LF$) derives from payment inflows ($f > 0$) exceeding their reserve requirements (Eq. \ref{eq: ib-s}). 
 \begin{align}
    & f_{i,t} = \left \{\begin{array}{ll}
      \sum_{i \in S_{i,t}} (C_{i,t} - W_{i,t})& \text{if} \; i = \text{commercial} \\
         \sum_{i \in S_{i,t}} (W_{i,t} - C_{i,t})  & \text{otherwise}
    \end{array} \right . \label{eq: flow}\\
    &DF_{i,t}= |f_{i,t}| - \Delta HPM_{i,t} \label{eq: ib-d}\\
    &LF_{i,t}= f_{i,t} - \Delta HPM_{i,t} \label{eq: ib-s}
\end{align}

\sloppy
\subsubsection{Interbank matching}\label{sec: matching}
The two banks can interact in the overnight and term segments of the unsecured interbank market. We model two interbank matching mechanisms, thus implementing two comparable scenarios. In the "Baseline" scenario, interbank matching depends on the amounts demanded and supplied, and lenders accommodate their customers' requests for funds and maturity. Instead, when the "Maturity" scenario is active, the interbank matching mechanism follows an \textit{active search for counterparties} \citep{Liu.2018} where borrowing banks search for lenders having the closest maturity preference. 

To do so, banks have target financial ratios that reflect their preferences for maturity which result from the maturity mismatch in their balance sheets.
While borrowing banks look at their maturity-based borrowing target ratios and decide how much to demand overnight and term, lending banks decide whether and how much to supply in the two segments according to their lending targets.
To model target ratios, this study exploits the Net Stable Funding Ratio (NSFR) of Basel III, a macroprudential tool that limits the excessive degree of banks' maturity mismatches characterizing the GFC. The NSFR requires that the total amount of \textit{available stable funds} (long-term liabilities $LM$) must at least equal the total amount of \textit{required stable funds} (long-term assets $AM$), that is $\sum_{i}a_{i}LM_{i}\geq\sum_{n}b_{n}AM_{n}$.
The concept of \textit{stability} revolves around the expiration dates of assets and liabilities. To do so, the weights $a_{i}$ and $b_{n}$ depend on the residual contractual maturities of the items on both sides of banks' balance sheets such that the longer the maturity, the higher the weight.\footnote{Liabilities are categorized according to their degree of stability, linked to their residual maturity, such that the higher the maturity, the higher the grade of stability, the higher the weight. Instead, assets are classified following their degree of liquidity such that the higher the maturity, the less the liquidity of the item, and the higher the weight.}
After some mathematical manipulations, we define a regulatory margin of stability $MS$ that drives banks' maturity choices in the interbank market.
According to the regulatory margin of stability, the amount of available stable funds must be at least equal to the amount of required stable funds, that is: 
\begin{equation}\label{eq:rms}\tag{a}
    \frac{a_{m}LM}{b_{m}AM}=MS_{i,t} \geq 1;
\end{equation}
where (i) $AM$ and $LM$ are unweighted assets and liabilities (Eq. \ref{eq: assets}-\ref{eq: liab}), (ii) $b_{m}$ is the fraction of required stable funds over total assets (Eq. \ref{eq: b_weight}), and (iii) $a_{m}$ is the proportion of available stable funds over total liabilities (Eq. \ref{eq: a_weight}).\footnote{The maturity weights follow the NSFR prescriptions; see Table \ref{tab:maturities}.} Please note that (i) interbank assets ($IBA$) and liabilities  ($IBL$) depend on the interbank exchanges that occurred in the overnight and term segments – defined in equations \ref{eq:onstock}-\ref{eq:termstock} – and that (ii) we distinguish between households' medium-term loans and firms' short-term credit when computing the proportion of required stable funds and assigning the corresponding maturity weight.
\begin{align}
   &AM_{i,t} =  L_{i,t-1}+HPM_{i,t-1}+B_{i,t-1}+B^{lr}_{i,t-1} + IBA^{on}_{i,t-1}+ IBA^{term}_{i,t-1}+R^{d}_{i,t-1} \label{eq: assets}\\
  & LM_{i, t} = D_{i,t-1} + IBL^{on}_{i,t-1}+ IBL^{term}_{i,t-1} + R^{l}_{i,t-1} + NPL_{i,t-1} + A_{i,t-1}\label{eq: liab}\\
    &  b_{i,t} = \left \{ \begin{array}{ll}
      \frac{\bar{m}_{1}IBA^{on}_{i,t-1} + \bar{m}_{2}(L_{i,t-1} + Bb_{i,t-1} + IBA^{term}_{i,t-1}) + \bar{m}_{3}B^{lr}_{i,t-1}}{AM_{i,t}}   & \text{if}\; i = \text{commercial}\\\\
     \frac{\bar{m}_{1}(L_{i,t-1} + IBA^{on}_{i,t-1}) + \bar{m}_{2}(Bb_{i,t-1}+IBA^{term}_{i,t-1})+\bar{m}_{3}B^{lr}_{i,t-1}}{AM_{i,t}}    & \text{otherwise}
    \end{array} \right . \label{eq: b_weight}\\
  & a_{i,t} = \frac{\bar{m}_{4}D_{i,t-1}+\bar{m}_{5}IBL^{term}_{i,-1}}{LM_{i,t}} \label{eq: a_weight} 
\end{align}

In the logic of this model, when the condition at Eq. \ref{eq:rms} is (not) satisfied, banks have a high (low) margin of stability and ask for overnight (term) funds on the interbank market, implying more (less) frequent market entries for funding purposes.
Target financial ratios are thus endogenous and must respect two conditions. As long as the target is not met, banks' demand for overnight funds must differ from zero. Conversely, banks' overnight requests must be null when the target is met.

To do so, banks demand in the overnight segment a \textit{proportion} of the total amount of reserves needed (Eq. \ref{eq:ibon}).  
\begin{align}
     &DF^{on}_{i,t}= DF_{i,t}(\theta_{t} \cdot \;\Pi^{b}_{i,t})\label{eq:ibon}
\end{align}
Overnight demand depends on two factors. First, a \textit{money market conditions parameter} $\theta_{t}$ that embodies banks' willingness to borrow overnight, dependent on interest rates profitability and the degree of perceived uncertainty (Eq. \ref{eq:theta}). 
\begin{align}\label{eq:theta}
\theta_{t}= \bar{a}_{0} + (\overline{icb}^{l}-i_{t-1}^{on}) + (i_{t-1}^{term}-i_{t-1}^{on})- \overline{PDU}; \; \text{with} \; \bar{a}_{0} \sim  U(0,1)
\end{align}

Second, a \textit{bank-specific parameter}, $\Pi^{b}_{i,t}\in[0,1]$, which represents banks' \textit{preferences for maturity} and incorporates the target ratios for overnight funds. This parameter computes the difference between the actual overnight borrowing ratio ($BOR_{i,t}$) and the targeted overnight borrowing ratios ($BOR_{i,t}^{T}$). The actual ratio is the proportion of "unstable" funding sources derived from the NSFR (Eq. \ref{eq:onb}). The target ratio is defined such that when banks have a very low degree of stability – when the regulatory margin of stability is not satisfied – their demand for overnight interbank is zero (Eq. \ref{eq:onbT}).
Otherwise, banks can afford overnight contracts since they satisfy the regulatory requirement, and their target ratio equals a random number $UN^{b}$ picked from the standard uniform distribution – $UN^{b}_{t} \sim U(0, BOR_{i,t})$ – which guarantees a positive overnight interbank demand.
\begin{align}
    &\Pi^{b}_{i,t}= BOR_{i,t}-BOR^{T}_{i,t}\\
    &BOR_{i,t}= 1 - a_{i,t} \label{eq:onb}\\
    &    BOR^{T}_{i,t}=
	\left\{ \begin{array}{llll}
    BOR_{i,t} &if& MS_{i,t}<1\\
		UN^{b}_{t} &if& MS_{i,t}\geq 1
		&\\	\end{array} \right.\label{eq:onbT}
\end{align}
When stability issues prevent banks from borrowing overnight, a null $\Pi^{b}_{i,t}$ ensures that the willingness to borrow overnight based on money market rates ($\theta_{t}$) does not play any role. Banks' access to the term segment of the theorized interbank market is residual (Eq. \ref{eq:ibterm}).
\begin{align}
    DF^{term}_{i,t} = DF_{i,t}-DF^{on}_{i,t}\label{eq:ibterm}
\end{align}

The formalization of the supply side is symmetrical. Overnight supply depends on the willingness to lend overnight and overnight lending ratios (Eq. \ref{eq:onl}).
\begin{align}
    LF^{on}_{i,t}= LF_{i,t}(L_{b}W_{t}\cdot \Pi^{l}_{i,t})\label{eq:onl}
\end{align}

The market parameter (Eq. \ref{eq:lbw}) depends on (i) the opportunity costs of providing funds on the overnight segment rather than on the term one or depositing free reserves at the central bank and (ii) the stress perceived on the interbank market $PDU$.
\begin{align}
    &L_{b}W_{t} = \bar{a}_{0} + \overline{PDU} + (i_{t-1}^{on}-\overline{icb}^{d}) -(i_{t-1}^{term}-i_{t-1}^{on}); \; \text{with} \; \bar{a}_{0} \sim  U(0,1)\label{eq:lbw}
\end{align}
Lending banks' preferences for maturity are captured by the difference between the actual overnight lending ratio ($LOR_{i,t}$) and the targeted overnight lending ratio ($LOR_{i,t}^{T}$).  The actual overnight lending ratio consists of the proportion of short-term assets that should not be backed by stable liabilities. 
About targeted ratios, the reasoning is symmetric to the demand side. A high degree of stability requires that either the amount of long-term liabilities must increase (numerator) or that the volume of long-term assets must decrease (denominator). When the degree of stability is low (high), long-term assets must decrease (increase): i.e., banks should lend more (less) overnight funds to align the expiration dates of assets and liabilities. Lending banks' supply in the term interbank segment is formalized residually (Eq. \ref{eq:ibterms}). In this case, $UN^{l}_{t}$ is bounded between zero and $LOR_{i,t}$.
\begin{align}
    & \Pi^{l}_{i,t} = LOR_{i,t}- LOR^{T}_{i,t}\\
    & LOR_{i,t} = 1 - b_{i,t}\\
     &LOR^{T}_{i,t}=
	\left\{ \begin{array}{llll}
        LOR_{i,t} &if& MS_{i,t}\geq 1 \\
		UN^{l}_{t} &if&MS_{i,t}<1\\
		\end{array} \right.\\
  &LF^{term}_{i,t}=LF_{i,t}-LF^{on}_{i,t}\label{eq:ibterms}
\end{align}

In the "Maturity" scenario, all the exchanges in the two segments of the interbank market thus depend on the values of (i) the market conditions parameters ($\theta_{t}$ and $L_{b}W_{t}$), and of (ii) the bank-specific parameters for maturities ($\Pi^{b}_{i,t}$ and $\Pi^{l}_{i,t}$).  Moreover, since lending banks might not accommodate borrowers' demand for funds, final interbank stocks reflect  the \textit{short-side} of the market (Eq. \ref{eq:onstock}-\ref{eq:termstock}). For any matching pairs of banks $i$ and $j$, such that $i$ is a lender and $j$ is a borrower and $\text{link}_{i,j}\neq 0$, interbank assets and liabilities are defined as follows.
\begin{align}
&IBA^{on}_{i, t} = IBL^{on}_{j, t} =
    \left\{ \begin{array}{lll}
     LF^{on}_{i, t}  & \text{if}\;& DF^{on}_{j, t}>LF^{on}_{i, t}\\
     DF^{on}_{j, t} & \text{if}\;& DF^{on}_{j, t}<LF^{on}_{i, t}
     \end{array} \right. \label{eq:onstock}\\
&IBA^{term}_{i, t} = IBL^{term}_{j,t}
    \left\{ \begin{array}{lll}
     LF^{term}_{i, t} & \text{if}\;& DF^{term}_{j,t}>LF^{term}_{i, t}\\
     DF^{term}_{j,t} & \text{if}\;& DF^{term}_{j,t}<LF^{term}_{i,t}
     \end{array} \right.  \label{eq:termstock}
\end{align}

Banks can access the central bank's standing facilities when interbank matching does not occur, whether it is amount- or maturity-based. While unmatched lending banks deposit their total loanable funds, matchless borrowers demand loans to the central bank for the full amount demanded in the interbank market; see equations \ref{eq:lendingfac}-\ref{eq:depfac}.
\begin{align}
& R^{l}_{i,t} = \left \{\begin{array}{ll}
 DF_{i,t} &  \text{if} \; \{j | j \in N_{b}\; \text{and} \; f_{j,t}>0\} = \emptyset \; \text{or} \; \text{link}_{i,j} = 0 \; \forall j \in N_{b} \; \text{s.t.}\; f_{j,t} > 0\\
   0  &  \text{otherwise}
\end{array} \right . \label{eq:lendingfac}\\
  & R^{d}_{i,t} = \left \{\begin{array}{ll}
 LF_{i,t} &  \text{if} \; \{j | j \in N_{b}\; \text{and} \; f_{j,t}<0\} = \emptyset \; \text{or} \; \text{link}_{i,j} = 0 \; \forall j \in N_{b} \; \text{s.t.}\; f_{j,t} < 0\\
   0  &  \text{otherwise}
\end{array} \right . \label{eq:depfac}
\end{align}

\begin{figure}[t]
    \centering
    \includegraphics[scale = 0.6]{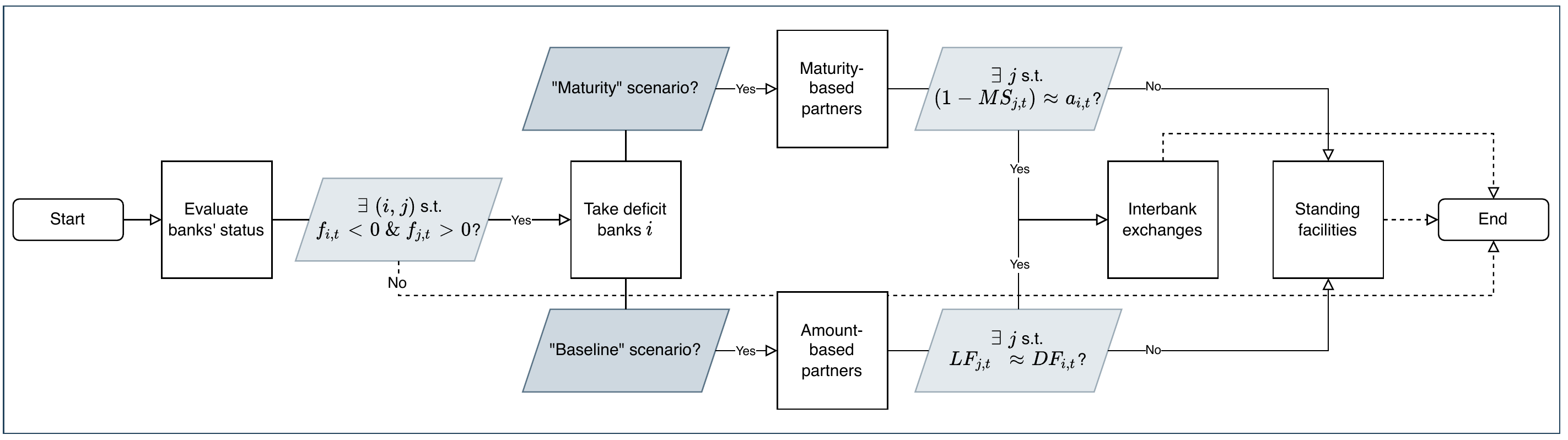}
    \caption{Interbank matching protocols.}
    \label{fig:figure3}
\end{figure}

Figure \ref{fig:figure3} summarizes the interbank matching protocols implemented.

\subsubsection{Interbank rates and funding costs} 
Interbank interest rates clear the overnight and term segments by adjusting to disequilibrium dynamics, i.e., excess supply or excess demand $\epsilon^{on}_{t}$ and $\epsilon^{term}_{t}$ \citep{reissl2018monetary, Reale.2022}.\footnote{These two measures are positive in case of excess demand and negative in case of excess supply.}
As such, the two rates must fall within the central bank's corridor.
\begin{align}\label{onint}
    &i_{t}^{on}=\overline{icb}^{d} +\frac{\overline{icb}^{l}-\overline{icb}^{d}}{1+e^{-(\bar{\sigma}_{ib}\epsilon^{on}_{t})}}\\
   &\epsilon^{on}_{t}= \sum_{j} DF^{on}_{j,t}-\sum_{i}LF^{on}_{i,t}
\end{align}
\begin{align}
   & i_{t}^{term}=\overline{icb}^{d} +\frac{\overline{icb}^{l}-\overline{icb}^{d}}{1+e^{-(\bar{\sigma}_{ib}\epsilon^{term}_{t})}}\\
   & \epsilon^{term}_{t}=\sum_{j}DF^{term}_{j,t}-\sum_{i}LF^{term}_{i,t}
\end{align}

Banks' funding costs (Eq. \ref{eq:fc}) are computed as the average between (i) the exogenous rate the central bank charges on required reserves ($\bar{icb}^{t}$), (ii) the interest rates borne on the interbank market, if exchanges occur, and (iii) the lending rate of the central bank's corridor if banks use the corresponding standing facility.\footnote{The central bank's policy rates, the deposit ($\overline{icb}^{d}$), and the lending rates ($\overline{icb}^{l}$) are exogenous and the corridor is assumed to be symmetric such that the targeted interest rate is defined as the average between the two, i.e. $\overline{icb}^{t}=\sfrac{(\overline{icb}^{l}+\overline{icb}^{d})}{2}$.} Please note that the denominator $x$ changes according to which trades actually occur at each simulation step.
\begin{align}
&\zeta_{i,t}=\frac{\overline{icb}^{t}+i^{on}_{t} + i^{term}_{t} + \overline{icb}^{l}}{x}\label{eq:fc}
\end{align}

Banks update credit market rates according to a mark-up/mark-down rule on previous period funding costs.

\begin{table}[t]
    \centering
   \resizebox{!}{0.1\paperwidth}{ \begin{tabular}{|lll|}
    \hline
     Policy experiments    &  Shocked variables & Step range\\\hline \hline
     Missing-shock & $–$ & $–$\\\hline \hline
      \multirow{2}{*}{Corridor-shock}   & $\overline{icb}^{l}  \overset{\mathrm{+}}{=} 50 \text{bp}$  & \multirow{2}{*}{$[300, 1200]\cap 300\mathbb{Z}$}\\ 
      & $\overline{icb}^{d}  \overset{\mathrm{+}}{=} 50 \text{bp}$ & \\ \hline \hline
      Width-shock &  $\overline{icb}^{l}  \overset{\mathrm{+}}{=} 50 \text{bp}$ & $[300, 1200]\cap 300\mathbb{Z}$ \\\hline \hline
      Uncertainty-shock & $PDU \overset{\mathrm{+}}{=}  0.2 $& $[300, 1200]\cap 300\mathbb{Z}$ \\ 
      \hline
    \end{tabular}}
    \caption{Policy experiments and shocks implemented.}
    \label{tab: shocks}
\end{table}
\section{Simulations and results}\label{section 4}

We run the model for $100$ parallel replicates over $1200$ steps and implement two scenarios depending on the interbank matching mechanism.\footnote{Stock-flow consistency checks are performed at the end of each simulation period. The model simulations are run on Julia, and the code supporting this study's findings is available from the corresponding author upon reasonable request.} In the "Baseline" scenario, banks are matched according to the total amounts requested and offered, demand and supply in the overnight segment only depend on money market parameters $\theta$ and $L_{b}W$, and no interbank rationing can occur as lenders accommodate their customer's requests for funds and maturity. When the "Maturity" scenario is active, banks' interbank matching mechanism depends on stability concerns – lenders' maturity mismatch ($1 - MS_{i,t}$) and borrowers' ASF ($a_{m}$) – derived from the NSFR construction explained in the previous section. 
The analysis conducted with this model aims to explore the implications of limiting exposures to excessive maturity mismatches and rollover risk on the clearing operations of the interbank market and also investigates the conditions under which mismatched maturity preferences between surplus and deficit banks affect the functioning of the interbank market and the conventional policies of central banks. To this aim, we perform three main policy experiments in both scenarios: (i) \textit{corridor-shock}, increasing the central bank's interest rates by $50$ basis points every $300$ steps, (ii) \textit{width-shock}, increasing the ceiling rate by $50$ basis points every $300$ periods while
keeping constant the floor rate, and (iii) \textit{uncertainty-shock}, gradually increasing the level of perceived uncertainty $PDU$ impacting the money market parameters $\theta$ and $L_{b}W$. Table \ref{tab: shocks} reports the policy experiments, the shocked variables, and the step ranges at which the shocks occur. For the sake of comparability, we plot the time series relative to the baseline setting where no experiment is conducted, i.e., \textit{missing-shock}. We report the results per each scenario and experiment – ruling out the first $100$ steps –  after having \textit{de-trended} the time series through the Hodrick-Prescott filter for monthly data.

\subsection{Policy experiments}
\subsubsection{Baseline scenario}
\begin{figure}[t]
\centering
\includegraphics[width = 0.8\paperwidth]{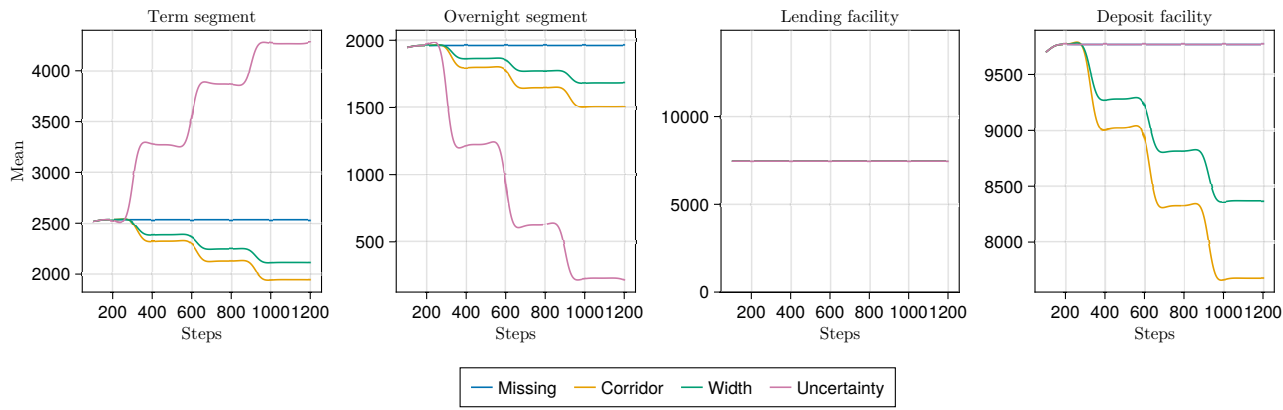}
\caption{Interbank market dynamics per shock in levels - Baseline scenario.}\label{fig:figure4}%
\end{figure}

Figure \ref{fig:figure4} depicts the dynamics of the interbank market for the "Baseline" scenario and each shock implemented in levels.
The "Missing-shock" time series (blue lines) serves as a benchmark to analyze the results of the experiments. Independently of the shock implemented, banks exchange more funds in the term interbank segment. However, the discrepancy between term and overnight funds is negligible when no shock occurs. An uncertainty shock exerts the highest impact on interbank dynamics: the higher the amount of perceived uncertainty, the greater the tendency of the overnight market to freeze. Indeed, when uncertainty reaches its highest peak ($0.8$) at the end of the simulation period, almost all funds are exchanged in the term interbank segment, whereas overnight funds tend to zero.
Rising uncertainty makes overnight funds decline by more than $80\%$, ultimately leading to a frozen overnight segment where the demand for overnight funds is almost null.
Since margins of stability are inactive and lenders are fully accommodating in this baseline setting, what matters most is the maturity at which deficit banks will be willing to borrow. Thus, this dynamic can be explained by $PDU$ negatively impacting borrowers' money market parameter $\theta$. This demand-led overnight freeze might not be harmful as long as the central bank has the ability to steer interbank rates through conventional monetary policies.\footnote{Appendix \ref{sec: appendixC} reports the evolution of interbank interest rates per each shock implemented. As we see from Fig. \ref{fig:figureC1}, the Baseline scenario allows the central bank to anchor interbank rates to its target, whatever the shock, since lending banks are always accommodating and no disequilibrium dynamics occur.}

\begin{figure}[t]
\centering
\includegraphics[width = 0.8\paperwidth]{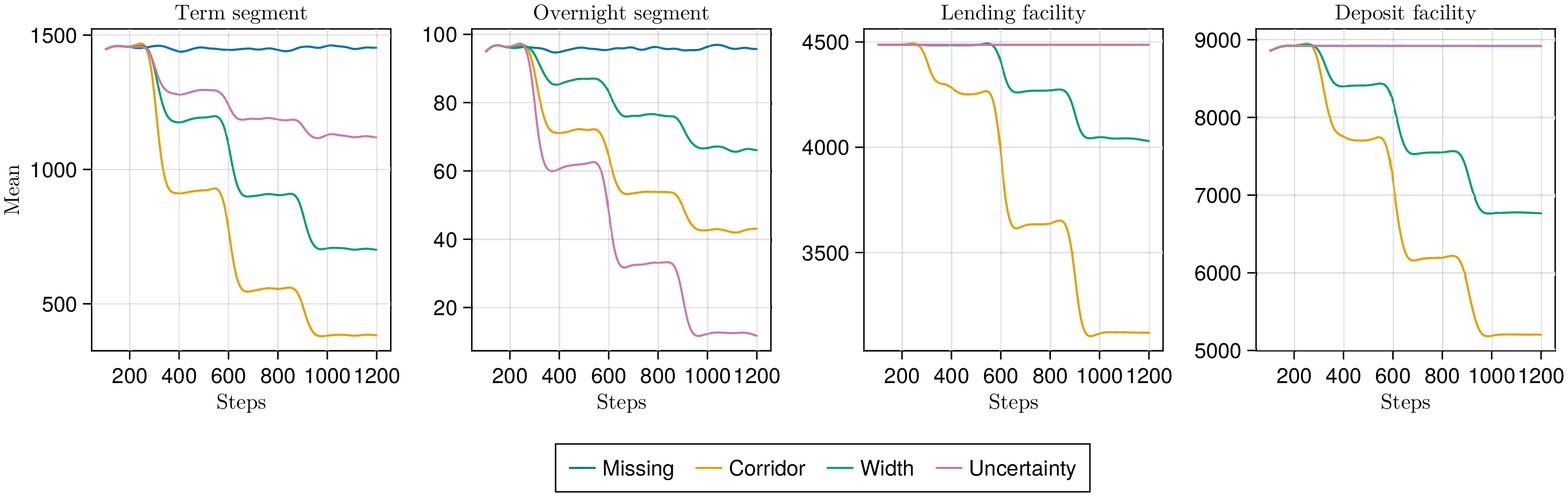}
\caption{Interbank market dynamics per shock in levels  - Maturity scenario.}\label{fig:figure5}%

\includegraphics[width = 0.5\paperwidth]{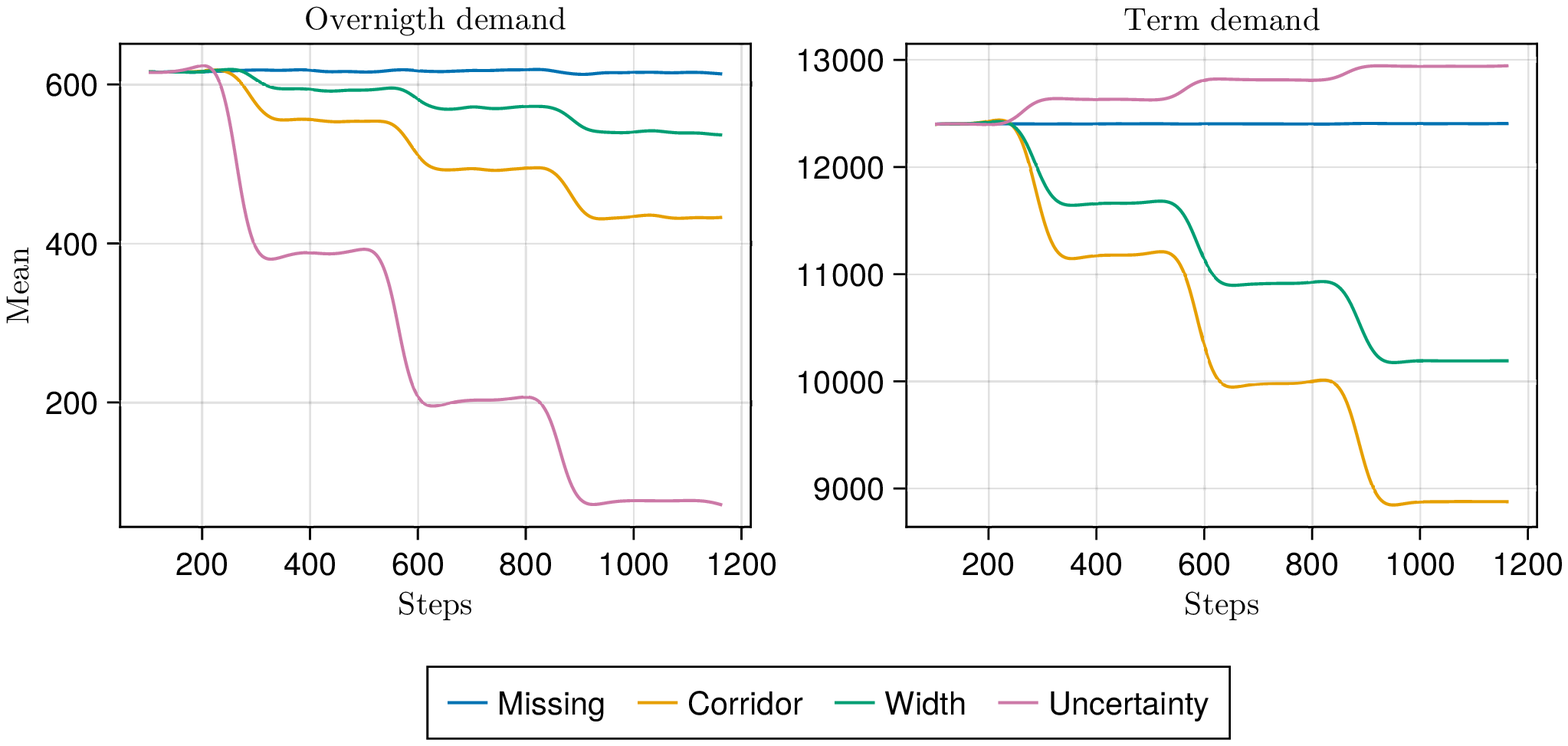}
\caption{Interbank demand in levels - Maturity scenario.}\label{fig:figure6}%
\end{figure}

The constant recourse to the lending facility reveals an effective matching mechanism throughout the simulation. As such, any bank needing reserves – after the initial burn-in period – can find a suitable partner in the interbank market who will accommodate its requests. Not only do banks look for lenders who can provide reserves in an amount that fits their deficit needs, but also flows net out in aggregate – as outflows and inflows are symmetrical (see Figure \ref{fig:figureC2} in Appendix \ref{sec: appendixC}) – making total demands and total supplies coincide. However, not all lenders can supply funds on the interbank market: unmatched surplus banks heavily use the central bank's deposit facility. The total amount of reserves deposited at the central bank almost reaches the overall volumes exchanged in both segments of the interbank market. Increasing policy rates – whether in the corridor or width shock – leads to a lower recourse to the deposit facility. This result might seem counterintuitive as a higher floor rate in the corridor shock should make lenders more inclined to deposit reserves rather than lending them in the money market. However, money market profitability does not impact banks' access to the deposit facility, and even if it did, $L_{b}W$ would be unaltered since interbank rates move along with corridor rates. Decreasing deposited reserves thus follows surplus banks' lower payment inflows, impacting their total loanable funds. 
As a result, increasing corridor rates weaken consumption expenditures and wage payments, reducing the amount of payment flows borne by banks and affecting the recourse to the deposit facility. An uncertainty shock, instead, does not pass through the real sector – as payment flows are unaltered – and only negatively impacts overnight exchanges. As such, without a well-functioning term interbank segment, banks' payments smoothing function would be compromised as uncertainty rises, even if supply interbank agents are fully accommodating and do not account for stability concerns.

\begin{figure}[t]
\centering
\includegraphics[width = 0.7\paperwidth]{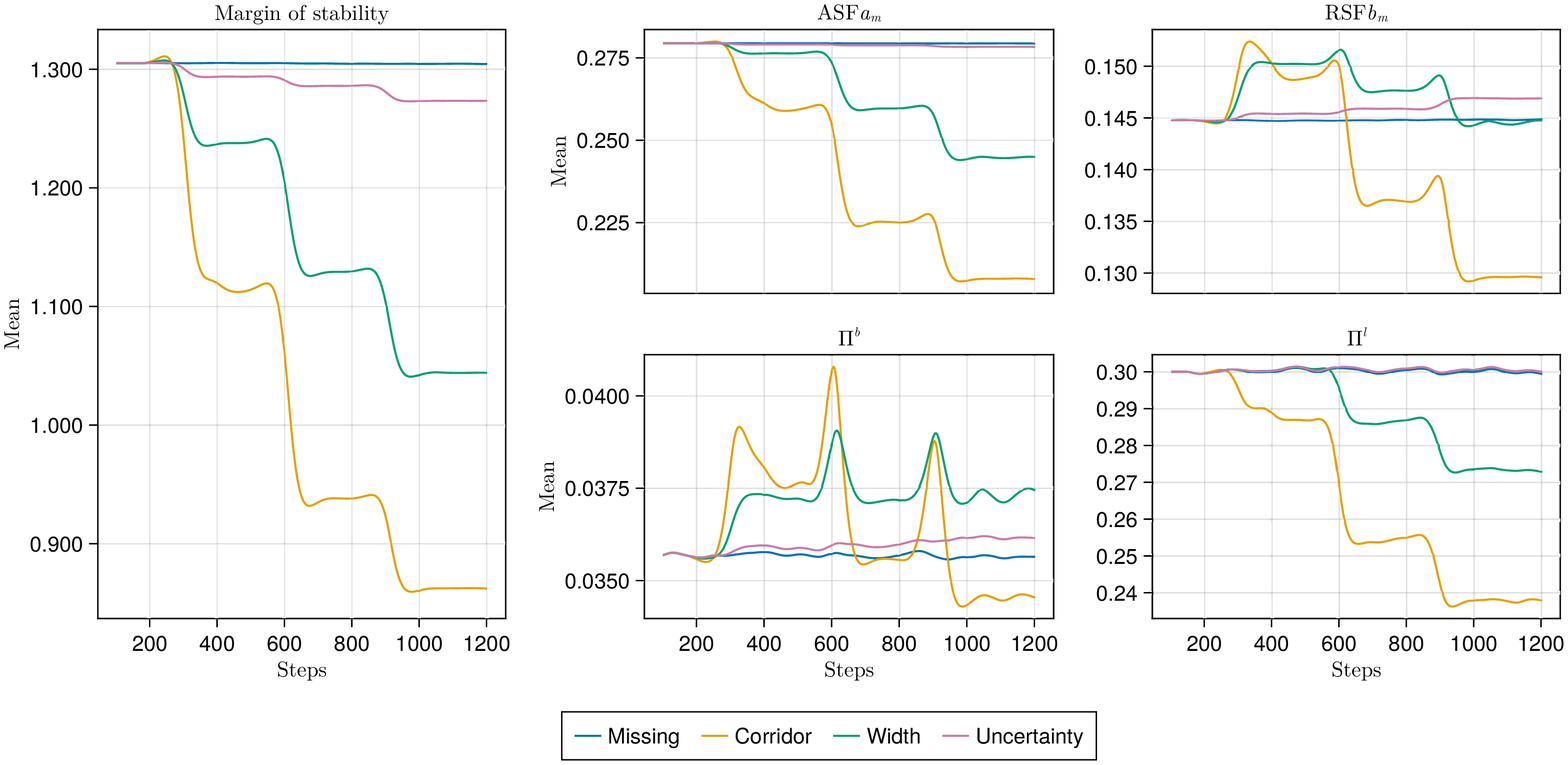}
\caption{Margin of stability and NSFR components in levels - Maturity scenario.}\label{fig:figure7}%

\includegraphics[width = 0.5\paperwidth]{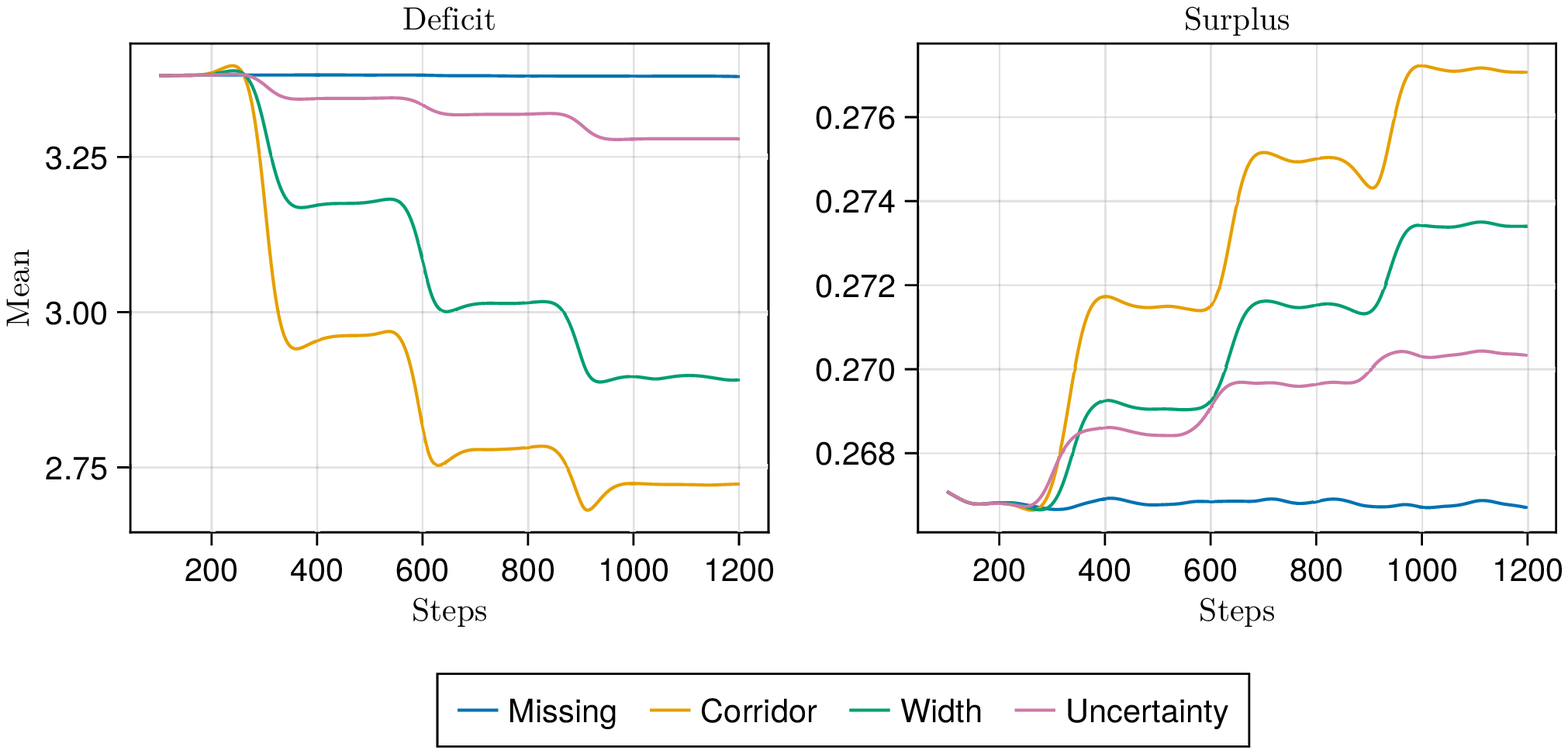}
\caption{Margin of stability diversified by interbank status - Maturity scenario.}\label{fig:figure8}
\end{figure}

\subsubsection{Maturity scenario}
\begin{figure}[t]
\centering
    \includegraphics[width = 0.8\paperwidth]{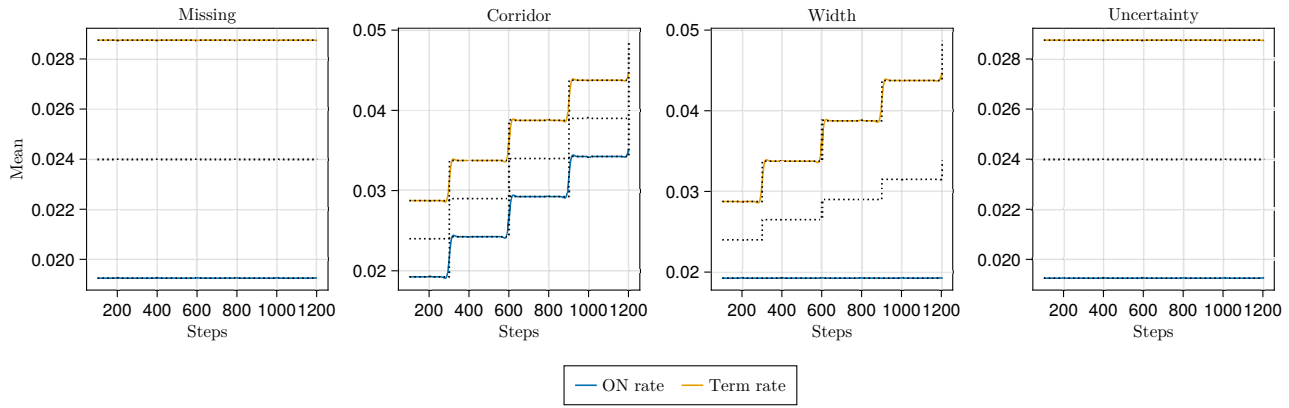}
\caption{Interbank rates - Maturity scenario.}\label{fig:figure9}%
\end{figure}

The results of the "Maturity" scenario are depicted in Fig. \ref{fig:figure5}. Interbank trades mainly occur in the term interbank market, with a large volume discrepancy between the two segments. The term segment is thus crucial to assess whether– and under which conditions – the whole money market may come to a standstill.  When maturity and stability issues drive banks' decision-making, the volumes exchanged in the overnight and term segments are respectively $95\%$ and $40\%$ lower than in the "Baseline" scenario (Fig. \ref{fig:figure4}). In other words, banks' rollover concerns may certainly lead the interbank market to freeze. 
Banks' usage of the central bank's facilities is quite high as the maturity-based interbank matching protocol compromises money market trades by making it more difficult for deficit banks to find a suitable partner. When uncertainty increases (pink line), the overnight segment tends to freeze even if $\Pi^{l}$ and $\Pi^{b}$ mediate the impact of money market parameters on short-term interbank demand and supply. Differently from what happens in the "Baseline" setting, rising $PDU$ negatively affects term trades, despite these decrease at lower rates than the ones exchanged in the overnight segment. This dynamic suggests the absence of a substitution effect between the two segments when the "Maturity" scenario is active. Indeed, borrowers' demand for term funds exhibits an increasing pattern (see Fig. \ref{fig:figure6}), which is, however, not accommodated by surplus institutions. The misalignment between maturity-based preferences is thus responsible for declining volumes in the interbank market as uncertainty grows. Uncertainty also indirectly impacts banks' stability (see Fig. \ref{fig:figure7}). While margins of stability, the proportion of stable funds over total liabilities (ASF), and target-based lenders' preferences for maturities ($\Pi^{l}$) slightly decrease, the amount of required stable funds over total assets (RSF) and borrowers' bank-specific parameter ($\Pi^{b}$) rise at a very low rate. Therefore, the impact of $PDU$ on NSFR components – and the derived overnight preferences – is negligible, and the stability of the overall banking sector seems to satisfy the requirements of the NSFR. However, the evolution of $MS_{i,t}$ diversified by interbank status (Fig. \ref{fig:figure8})
suggests that deficit (surplus) banks experience an overall high (low) degree of stability. This dynamic is not surprising as the higher amount of term liabilities makes borrowers' ASF increase (the numerator of NSFR). Conversely, surplus banks' long-term assets make $b_{m}$ rise (the denominator of NSFR), discouraging stability. 

When the system is shocked by increasing corridor rates (orange and green lines), the drop in interbank exchanges and standing facilities is stronger when a corridor shock is implemented. However, borrowers' access to the lending facility starts to decrease only after the second round of a width-shock at around step $600$ (see Fig. \ref{fig:figure5}), i.e., after the corridor width widens by $100$ bp. This dynamic is led by the evolution of $\Pi^{l}$ – Fig. \ref{fig:figure7} – which responds to a width-shock with a $600$ steps lead and indicates that deteriorating lending overnight targets indirectly feedback into the matching protocol, altering the volumes of interbank assets and liabilities which impact lenders' maturity mismatch and borrowers' stability. Overall, altering the width makes the system more resilient in the short term as banks' portfolios are either temporarily (RSF) or moderately (ASF) affected when the shock starts at step $300$. A width shock has thus a lower impact on banks' balance sheet composition and the consequent NSFR-based risk factors $a_{m}$ and $b_{m}$. Moreover, the higher the width, the higher the tendency of required stable funds (RSF) to converge towards steady state dynamics (blue line). This explains why stability margins do not fully drop as in the corridor shock. 
Conversely, a symmetric increase in corridor rates – corridor-shock – strongly deteriorates the stability of the overall banking system (Fig. \ref{fig:figure7}), eventually leading to unsatisfied NSFR in the long run.
The impact of policy rate shocks on banks' stability margins is also asymmetric. When we look at deficit vs. surplus banks (Fig. \ref{fig:figure8}), higher rates weaken deficit banks' $MS_{i,t}$ more than they strengthen lenders' stability conditions. It is therefore important to pay special attention to deficit banks, whose decreasing stability could ultimately lead to unsatisfied NSFRs and further frictions on the interbank market due to lenders' refusal to meet the maturities of their customers. Maturity-based issues become relevant in this scenario as the central bank can not steer interbank interest rates toward the desired target through conventional monetary policies, see Fig. \ref{fig:figure9}. Divergent maturity preferences between lenders and borrowers in the money market thus lead to disequilibrium dynamics that make the overnight rate (term rate) to be anchored to the policy floor (ceiling) due to a constant excess supply (demand) condition.

\begin{figure}[t]
    \centering
    \includegraphics[width = 0.5\paperwidth]{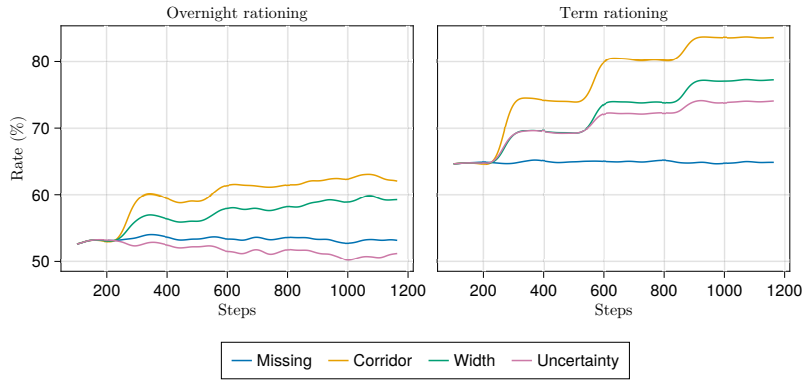}
\caption{Interbank rationing per shock - Maturity scenario.}\label{fig:figure10}
\end{figure}

The "Maturity" scenario is characterized by non-accommodating lenders who can ration their clients due to maturity preferences. Interbank rationing thus endogenously emerges from agents' interactions and the evolution of surplus and deficit banks' degree of stability. Fig. \ref{fig:figure10} depicts the evolution of a rationing measure ($\Gamma$) for the overnight and term segments. $\Gamma$ has been constructed such that rationing is null when the amount of liabilities for a certain market ($IBL^{x}$) equals demand ($DF^{x}$): $\Gamma^{x} = (1 - \frac{IBL^{x}}{ DF^{x}}) \in [0,1]$, for any generic sub-market $x = \{\text{ON}, \text{Term}\}$.
In the absence of shocks (blue line), roughly $55\%$ of overnight customers are rationed, compared to the $65\%$ in the term segment. 
The shocks impact the overnight segment asymmetrically: while uncertainty (slightly) reduces the amount of unsatisfied customers, policy rate shocks exert the opposite effect. Despite a decreasing demand (Fig. \ref{fig:figure6}) should lead rationing to shrink, policy rate shocks 
trigger a higher proportion of rationed interbank customers. In this case, final interbank liabilities in the corridor- and width-shocks seem to follow decreasing lenders' supplies driven by $\Pi^{l}$, making both $IBL$ and $DF$ decrease. 
Term rationing almost reaches $100\%$ under a corridor shock since term trades are highly discouraged by synchronized increases in policy rates. Therefore, (i) a corridor-based contractionary shock (orange line) combined with surplus banks highly exposed to rollover risks may result in severe frictions in the term segment, while (ii) a width-based contractionary monetary policy shock (green line) does not fully compromise the efficient functioning of the interbank market.

\begin{figure}[t]
\centering
\includegraphics[width = 0.5\paperwidth]{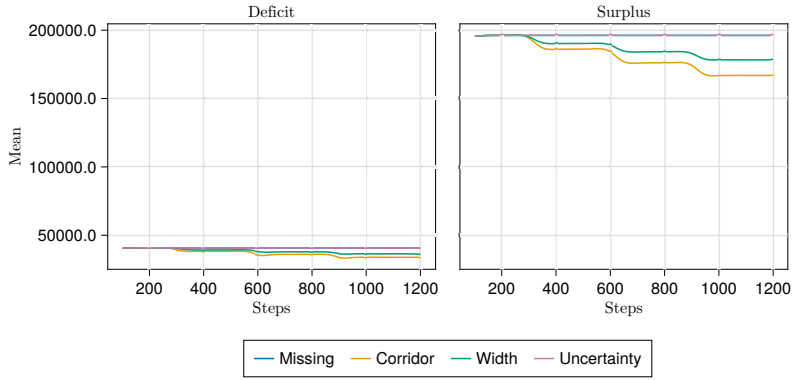}
\caption{Real sector loans  - Maturity scenario.}\label{fig:figure11}
\end{figure}

Stability divergences between surplus and deficit banks become relevant within a fragmented monetary union, like the European one, characterized by a clear distinction between reserve-scarce peripheral banks and surplus-oriented core banks. During the European interbank fragmentation, deficit banks in the periphery of the Euro Area faced higher funding costs to borrow in the interbank market, discouraging credit provision \citep{Berkmen.2013}. This dynamic clearly emerges in this theorized framework:  Fig. \ref{fig:figure11} shows that loans granted by deficit banks are constantly lower in volume than those provided by surplus banks.\\

Moreover, European core surplus banks' refusal to accommodate interbank borrowers' requests triggered 
asymmetric funding barriers and led to a halt of overnight exchanges, compromising the central bank's actions \citep{Tropeano.2019} and revealing country-specific funding barriers \citep{Bonatti.2021} which are incompatible with the conduct of a monetary union \citep{Lapavitsas.2010, Mayordomo.2015}.
Despite T2 \textit{imbalances accumulation} being outside this model's scope – as we model symmetric payment flows happening within the same country – the "Maturity" scenario tends to replicate some interesting dynamics of the recent interbank fragmentation that led to a massive recourse to the standing facilities and forced the European central bank to intervene with unconventional measures.  As such, the asymmetric funding frictions behind the interbank fragmentation could also be induced by mismatched preferences for maturities and the unbalanced distribution of rollover risks within the banking sector. While these considerations point to an additional rollover-based source of interbank segmentation that goes beyond the usual sovereign- \citep{Cesaratto.2013, DeSantis.2019} and credit-risk \citep{Eisenschmidt.2018} dichotomy, we leave the interaction between different types of risks and a multi-country analysis for future research. However, these results are in line with \cite{Bechtel.2019}'s empirical findings who emphasize that (i) immediate settlements of real-sector transactions exacerbate banks’ liquidity management and can be an additional source of financial instability, and (ii) that
rollover-based risks triggering banks' heterogeneous funding costs may also persist in periods of \textit{financial quietness}, i.e., when the banking sector's stability is not jeopardized.

\subsection{Sensitivity analysis}

We  perform sensitivity tests to check how altering the weights of the NSFR impacts the steady-state dynamics of the "Maturity" scenario (missing shock). To do so, we alter the parameters characterizing the computation of $a_{m}$ and $b_{m}$ (Eq. \ref{eq: a_weight} - \ref{eq: b_weight}) from $0$ to $1$ incremented by $0.1$. Please note that the higher variability of the time series compared to the figures presented in the previous section derives from the single-run simulation performed during this sensitivity analysis. We thus plot a 200-step moving average to gain a clearer understanding of the underlying dynamics.
Table \ref{tab:maturities} summarizes the parameters under investigation, the corresponding variables, and the theoretical maturity assigned following the NSFR \citep{NSFR.2014}. The figures reporting the analysis for $m_{2}$ and $m_{5}$ are in Appendix \ref{sec: appendixC}, as no significant alteration to the baseline trend occurs. Each test is conducted by altering one parameter at a time, all else being equal. 

\begin{table}[H]
\centering
\resizebox{0.6\textwidth}{!}{\begin{tabular}{|l|lllll|}
\hline 
&Parameter &Variable  & Maturity & NSFR-value & Tests range\\ 
\hline \hline
\multirow{3}{*}{Assets} & $m_{1}$&  $Lf$, $IBA^{on}$ & $M<6m$ & $0.1$ &$[0, 1] \cap 0.1 \mathbb{Z}$ \\
& $m_{2}$& $Bb$, $Lh$, $IBA^{term}$ & $6m \leq M \leq 1y$ & $0.5$  &$[0, 1] \cap 0.1 \mathbb{Z}$\\
&$m_{3}$ & $B^{lr}$&$M\geq 1y$&$0.05$  &$[0, 1] \cap 0.1 \mathbb{Z}$\\
\hline \hline

\multirow{2}{*}{Liabilities} & $m_{4}$ & $Dh$, $Df$ & & $0.9$  &$[0, 1] \cap 0.1 \mathbb{Z}$\\
 &$m_{5}$& $IBL^{term}$ & $6m \leq M \leq 1y$ & $0.5$  &$[0, 1] \cap 0.1 \mathbb{Z}$\\
\hline
\end{tabular}}
\caption{Implicit contractual maturities and weights; $m$ stands for months and $y$ for years.}\label{tab:maturities}
\end{table}

\begin{figure}[H]
    \centering
    \includegraphics[width = 0.8\paperwidth]{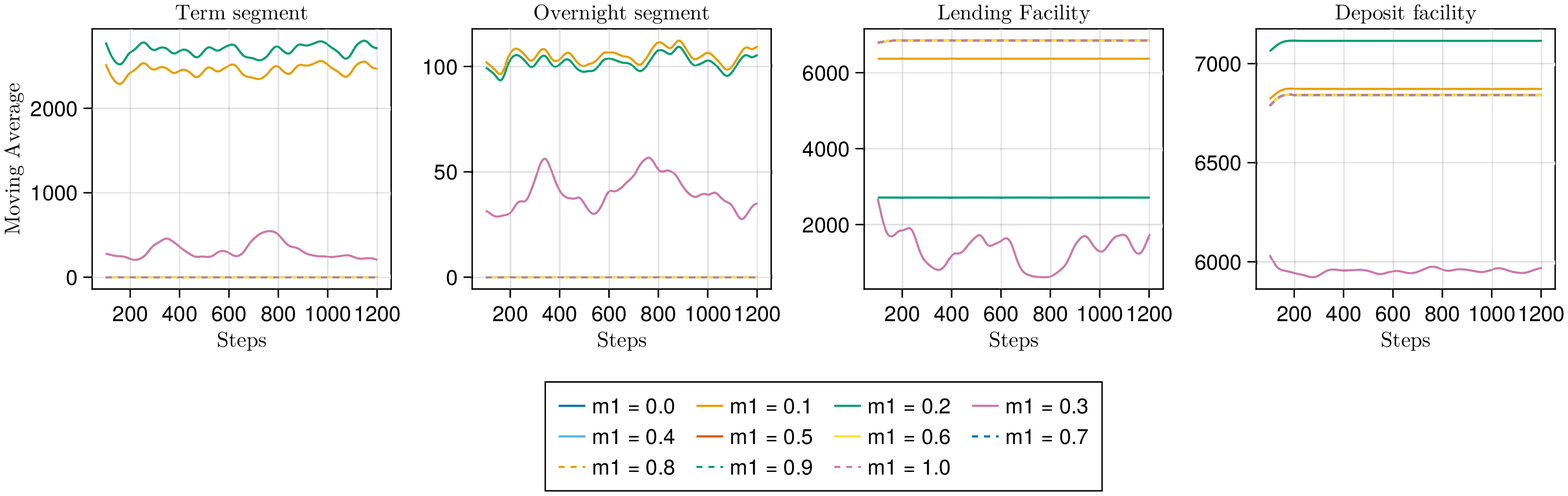}
    \caption{Sensitivity analysis over NSFR parameter $m_{1}$. The baseline value used in the main simulations corresponds to $m_{1} = 0.1$ (orange line).}
    \label{fig:figure12} %
    
            \includegraphics[width = 0.8\paperwidth]{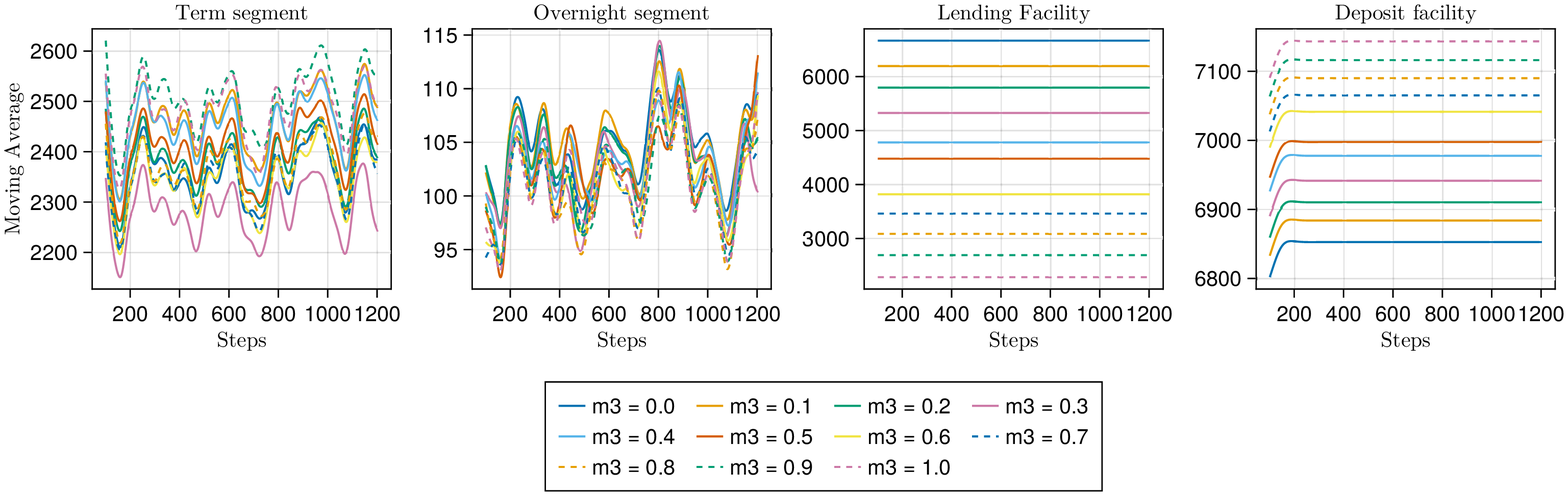}
    \caption{Sensitivity analysis over NSFR parameter $m_{3}$. The baseline value used in the main simulations corresponds to $m_{3} = 0.05$.} \label{fig:figure13}%
\end{figure}

Fig. \ref{fig:figure12} reports the results of the analysis conducted on $m_{1}$, which impacts the RSF factor of short-term firms' loans and overnight interbank volumes.  Looking at the volumes exchanged in the interbank market, all parameter values outside the range $[0.1, 0.3]$ lead to a full freeze of both segments. Indeed, a high weight assigned to short-term loans strongly alters the ability of reserve-scarce banks to find lenders in the interbank market, making the banking sector more exposed to the central bank's standing facilities. When $m_{1}$ is $0.2$, term (overnight) trades are slightly higher (lower). This symmetric effect arises because a slight increase in $m_{1}$ induces $b_{m}$ to rise – the denominator of the NSFR – hindering banks' stability. Consequently, banks look for more stable portfolio configurations and are more prone to hold term assets and liabilities. However, further increases in this short-term RSF factor may have negative effects on the whole system, as frozen interbank activities compromise the functioning of the payment system and the theorized economy as a whole. This dynamic could be interpreted as a sign of increasing holdings of illiquid assets that could not be easily rolled over or sold at fire sales prices.

\begin{figure}[t]
    \centering
     \includegraphics[width = 0.5\paperwidth]{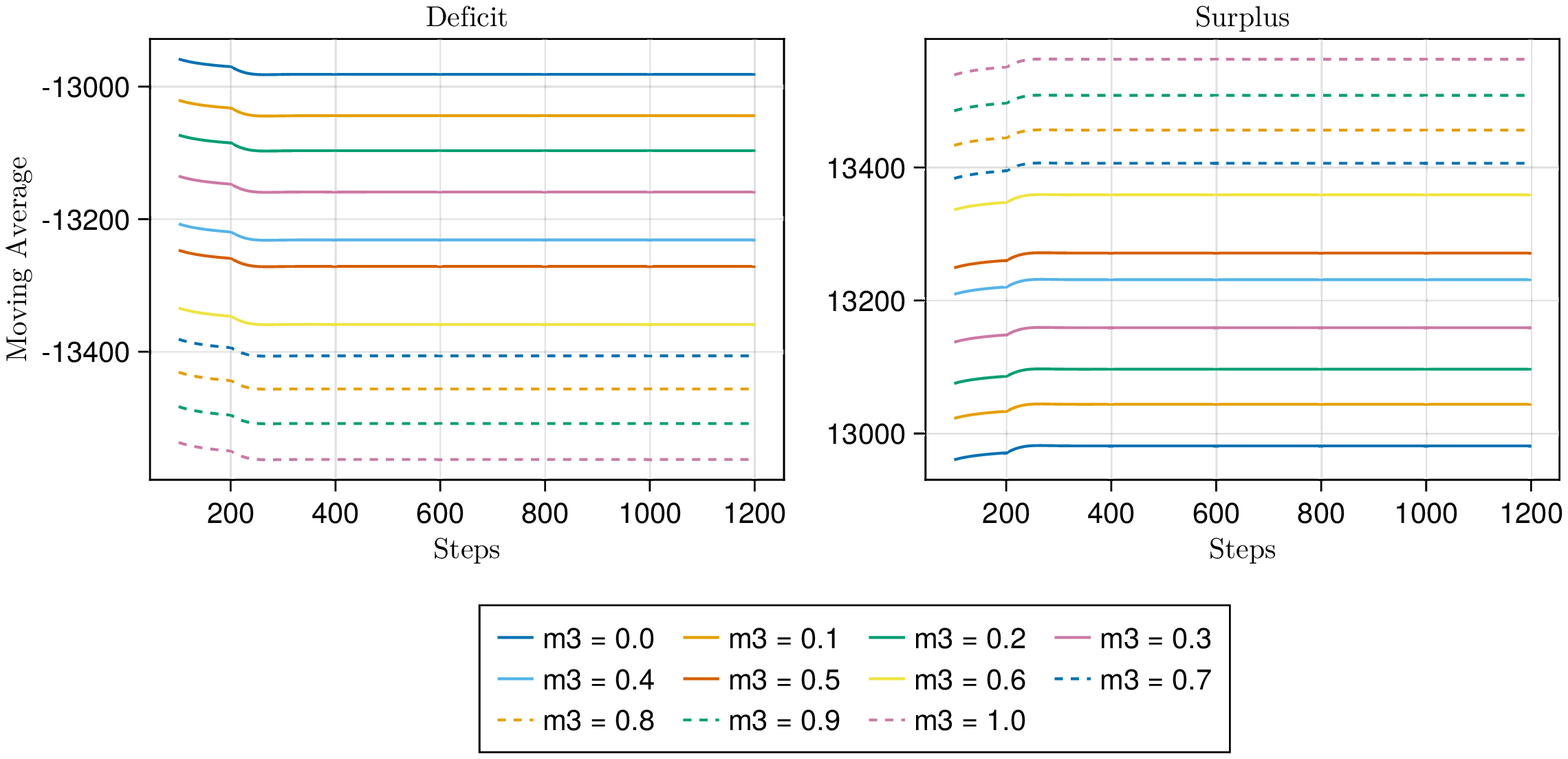}
    \caption{Payment flows evolution as $m_{3}$ changes.}
    \label{fig:figure14} %

           \includegraphics[width = 0.8\paperwidth]{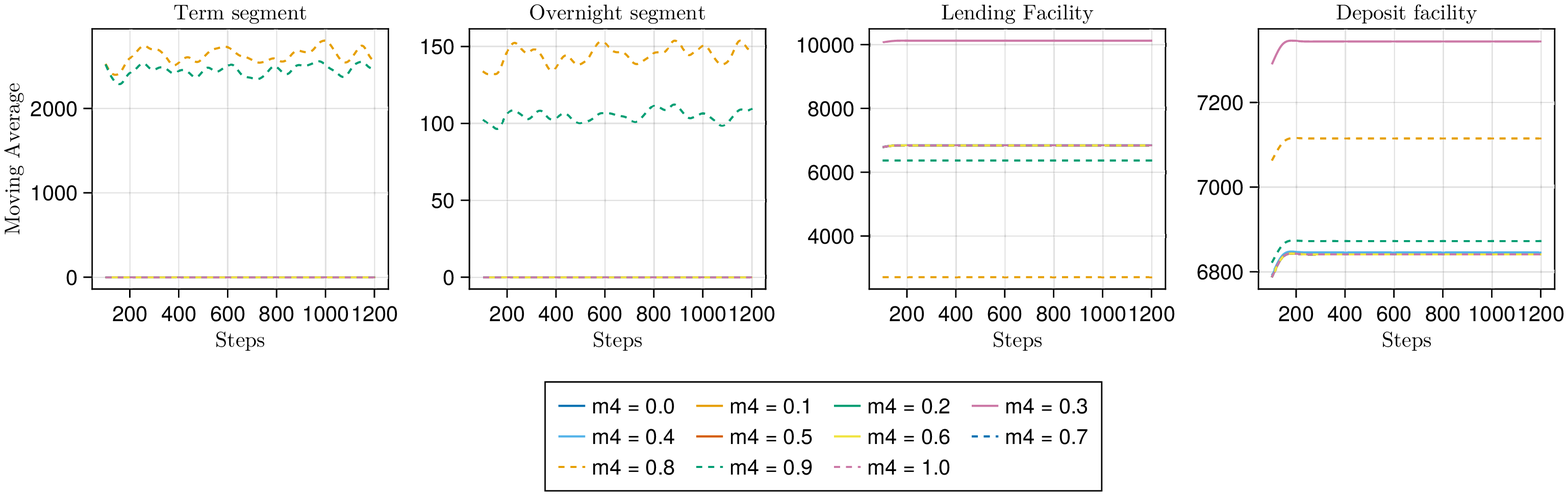}
    \caption{Sensitivity analysis over NSFR parameter $m_{4}$. The baseline value used in the main simulations corresponds to $m_{4} = 0.9$ (dashed green line).}
    \label{fig:figure15}
\end{figure}

Variations of the parameter $m_{3}$ (Fig. \ref{fig:figure13}), i.e. the weight assigned to long-term government bond holdings, do not substantially affect the volumes exchanged in the interbank market as the curves overlap despite the higher variability. The impact on standing facilities is, however, linear and symmetric. Higher values of $m_{3}$ make borrowing banks less exposed to the central bank's lending facility and surplus ones more willing to deposit reserves at the monetary authority's account. Since interbank trades are not affected, the recourse to the standing facilities does not change because of the matching protocol. Instead, it is due to how $m_{3}$ impacts the evolution of payment flows of surplus and deficit banks, thus altering total demand and supply, see Fig. \ref{fig:figure14}. Altering the weight of government bonds thus indirectly affects the levels of key variables of the theorized economy – though not influencing the underlying dynamics – despite we do not account for securitized interbank segments where $B^{lr}$ could be used as collateral. The banking system thus appears resilient to changes in $m_{3}$, although changes in the NSFR parameters pass through the real economy.

A slight decrease of the weight assigned to households' and firms' deposits ($m_{4}$) from $0.9$ (baseline) to $0.8$ (dashed orange line) – see Fig. \ref{fig:figure15} – boosts both interbank volumes and banks' access to the standing facilities, while leaving unaffected the higher reliance on term exchanges. This low decrease in $m_{4}$ should, however, discourage overnight loans as a higher ASF risk factor reduces banks' margins of stability.  This dynamic does not emerge because stability margins are overall satisfied for values of $m_{4}$ closer to the baseline setting.
Instead, as the parameter is lower than $0.5$, the stability of the banking sector as a whole is strongly compromised, leading to a halt of bilateral interbank trades and a stronger usage of both policy facilities. 
Indeed, a lower $m_{4}$ implies that term deposits lose stability as the propensity of customers to withdraw their funding increases, jeopardizing financial stability.

Alterations of short-term loans and deposit weights ($m_{1}$ and $m_{4}$ respectively) thus exert the highest impact on the model's dynamics. Some parameter values could also lead the whole interbank market to freeze, boost the recourse to the standing facilities, and compromise the maturity-based stability of the banking system. As such, policymakers and research practitioners should not neglect the interaction between the NSFR and rollover-based interbank dynamics. This issue is of greater importance within the European context, where asymmetric funding barriers may also be associated with unbalanced stability conditions and mismatched maturity preferences between peripheral deficit banks and core surplus institutions.

\section{Conclusions}\label{section 5}
This paper developed a complex Agent-Based Stock-Flow Consistent analysis of a potential payment system within an overdraft economy where two collections of heterogeneous banks can interact in two segments of the unsecured interbank market diversified by maturity, overnight, and term. We differentiate between commercial banks providing loans only to households and business banks granting credit only to firms to finance production. By doing so, demand and supply for interbank funds originate from payment settlement purposes to smooth payment liquidity shocks in the form of inflows and outflows.
In this study, we investigate how rollover-based interbank decisions may impact the well-functioning of the money market and eventually compromise the central bank's conventional measures.
To do so, we model two interbank matching scenarios. In the "Baseline" scenario, banks look for those lending partners who can best accommodate their demand for overnight and term funds. Conversely, the "Maturity" scenario accounts for potential interbank rationing as borrowers' and lenders' decisions are driven by bank-specific maturity targets, which depend on the dictates of the NSFR imposed by Basel III.
When matching is unsuccessful, banks can rely on the central bank's standing facilities.

We shock both scenarios by (i) altering the degree of perceived uncertainty and (ii) imposing a contractionary monetary policy shock which can take the form of a symmetric corridor shock or an asymmetric width shock. As uncertainty rises, overnight interbank trades in the "Baseline" scenario come to a standstill and are substituted by higher volumes exchanged in the term unsecured segment. As a result, banks' ability to smooth payments efficiently depends on a well-functioning term market, even though surplus banks can accommodate all interbank requests.
This substitution effect is, however, absent when we introduce non-accommodating lenders and maturity-based interbank preferences. Maturity misalignment between deficit and surplus banks thus compromises the well-functioning of the interbank market, boosts the recourse to the monetary authority's standing facilities, and makes monetary policy interest-rate steering practices ineffective. Moreover, the results show a clear divergence between surplus and deficit banks' degrees of stability. While borrowing banks are overall more stable, due to the high reliance on term funds, lending institutions find themselves more exposed to maturity mismatch and rollover risk, which, however, does not influence the overall stability of the banking sector. This dual stability-based configuration resembles the segmented European interbank structure with reserve-scarce peripheral banks vs surplus core ones. The results of the "Maturity" scenario, indeed, capture some stylized facts of the European fragmentation, despite T2 imbalances accumulation being outside this model's scope. As such, the simulations point to the potential emergence of a rollover-induced interbank fragmentation, which could anticipate the usual credit- and sovereign-risk explanations. The heterogeneous exposure to rollover-based risks and maturity mismatches may thus lead to asymmetric funding frictions that persist despite the high banking sector's stability and (dis-)encourage surplus (deficit) institutions' provision of credit.
The sensitivity analysis on the weights of the NSFR shows that alterations to the liquidity of short-term loans and the stability of deposits may compromise the whole functioning of the interbank market and thus expose the theorized system to severe financial instability. As such, a joint analysis of the NSFR and rollover-based interbank dynamics could be useful in understanding how unbalanced stability conditions of heterogeneous banking sectors respond to regulatory changes. 

This study is the first step in a wider research agenda. The aim of future works is to extend this framework to a multi-country AB-SFC setting with fully heterogeneous sectors and institutions. To enhance the model's ability to resemble the functioning of the T2, this framework should also incorporate the interaction between various interbank segments diversified by maturity and the degree of collateralization.
Several other aspects of this work should be addressed more in the future, including default mechanisms, non-accommodating lenders in the credit market, heterogeneous behaviors of agents in the real sector, and the interaction between credit, sovereign, and funding liquidity risks.

\newpage
\begin{appendices}
\setcounter{table}{0}
\setcounter{figure}{0}
\renewcommand{\thetable}{\Alph{section}\arabic{table}}
\renewcommand{\thefigure}{\Alph{section}\arabic{figure}}

\section{Additional figures} \label{sec: appendixC}

\begin{figure}[H]
    \centering
    \includegraphics[width = 0.8\paperwidth]{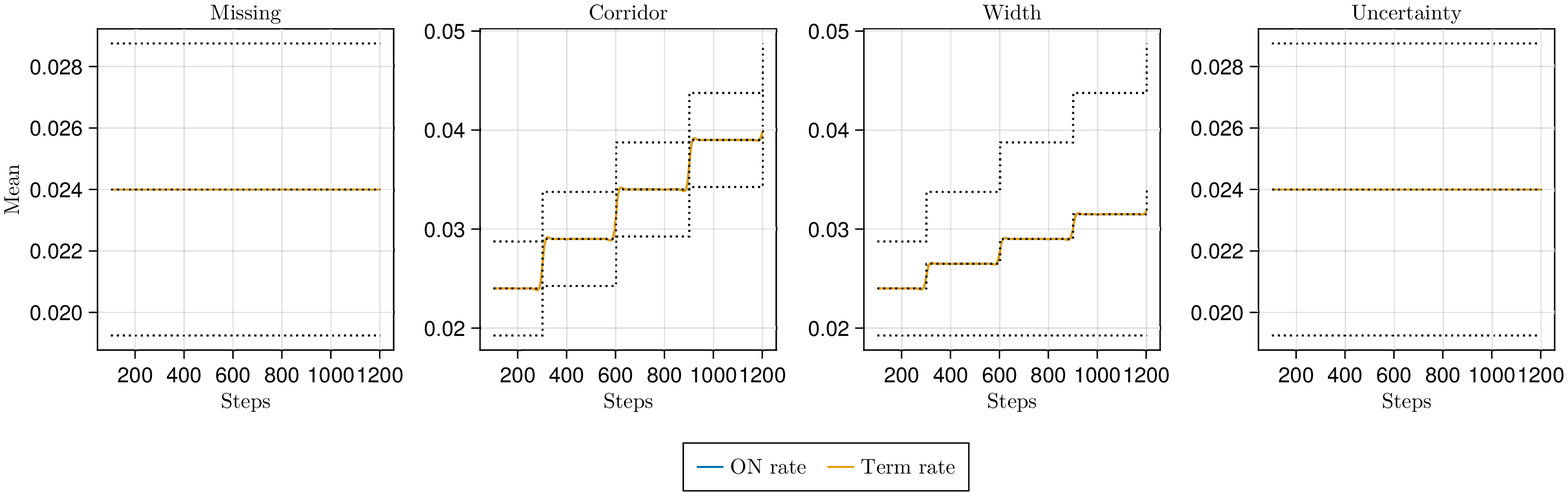}
    \caption{Interbank interest rates per shock – Baseline scenario.}
    \label{fig:figureC1}

    \includegraphics[width = 0.8\paperwidth]{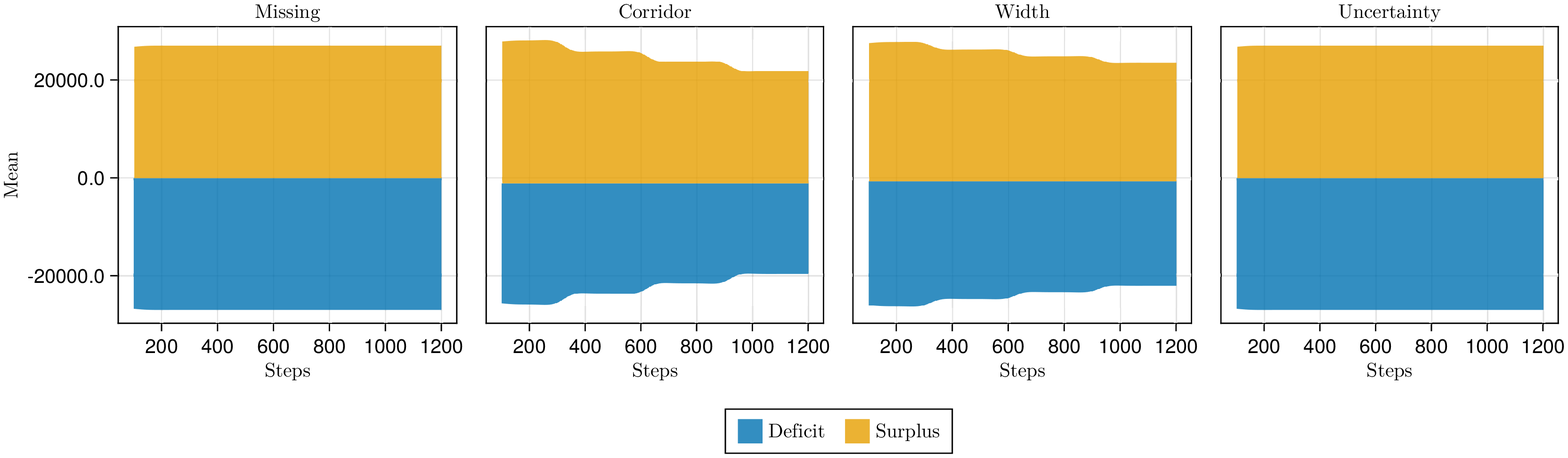}
    \caption{Payment flows balances per shock – Baseline scenario.}
    \label{fig:figureC2}
\end{figure}

\begin{figure}[H]
    \centering
        \includegraphics[width = 0.8\paperwidth]{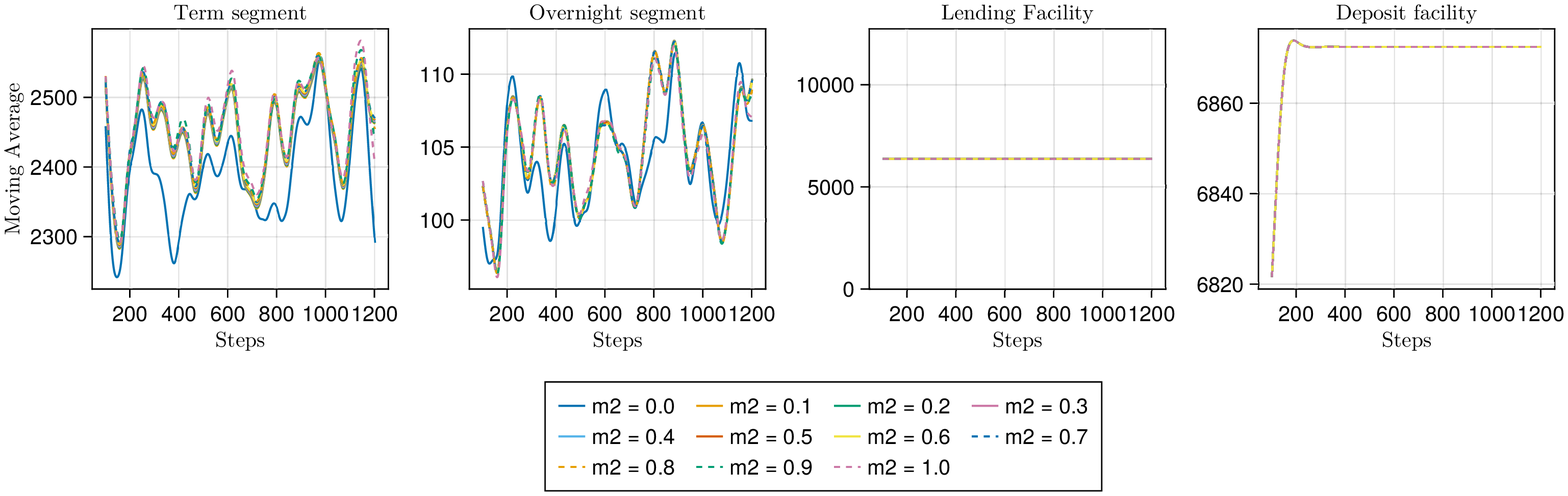}
    \caption{Sensitivity analysis over NSFR parameter $m_{2}$. The baseline value used in the main simulations corresponds to $m_{2} = 0.5$ (red line).}
    \label{fig:figureC3}
\end{figure}

\begin{figure}[H]
    \centering
        \includegraphics[width = 0.8\paperwidth]{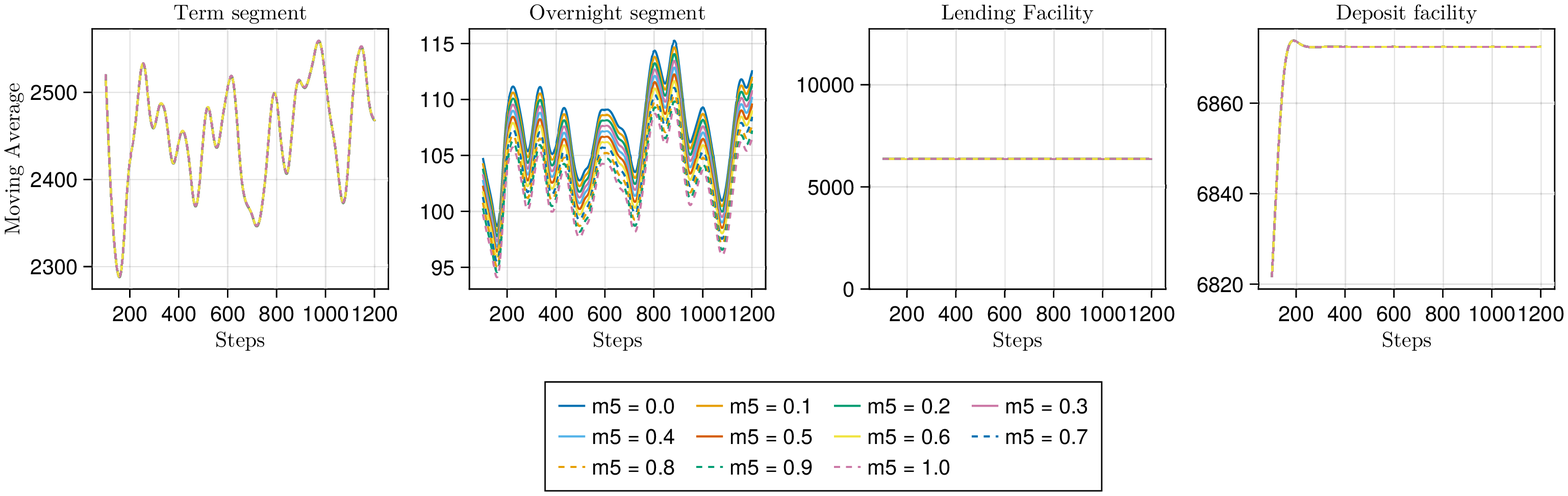}
    \caption{Sensitivity analysis over NSFR parameter $m_{5}$. The baseline value used in the main simulations corresponds to $m_{5} = 0.5$ (red line).}
    \label{fig:figureC4}
\end{figure}
\newpage

\section{Additional sensitivity analyses}\label{sec: appendixD}
\setcounter{figure}{0}
In addition to the sensitivity tests performed on NSFR parameters, we perform robustness checks on our initial assumptions concerning the debt-to-GDP ratio ($r$), capital depreciation ($\delta$), the share of non-performing loans ($l$), households' leverage ($\gamma$), and firms' propensity for deposits ($g_{d}$). We run the analysis for the "Baseline" scenario for several parameter ranges. We report the results for firms' output and  the overnight interbank segment in Fig. \ref{fig:figureD1}-\ref{fig:figureD2}. The results show that varying levels of the parameters of interest within the ranges considered do not alter the system's dynamics, as only variable levels are affected.

\begin{figure}[H]
    \centering
        \includegraphics[width = 0.7\paperwidth]{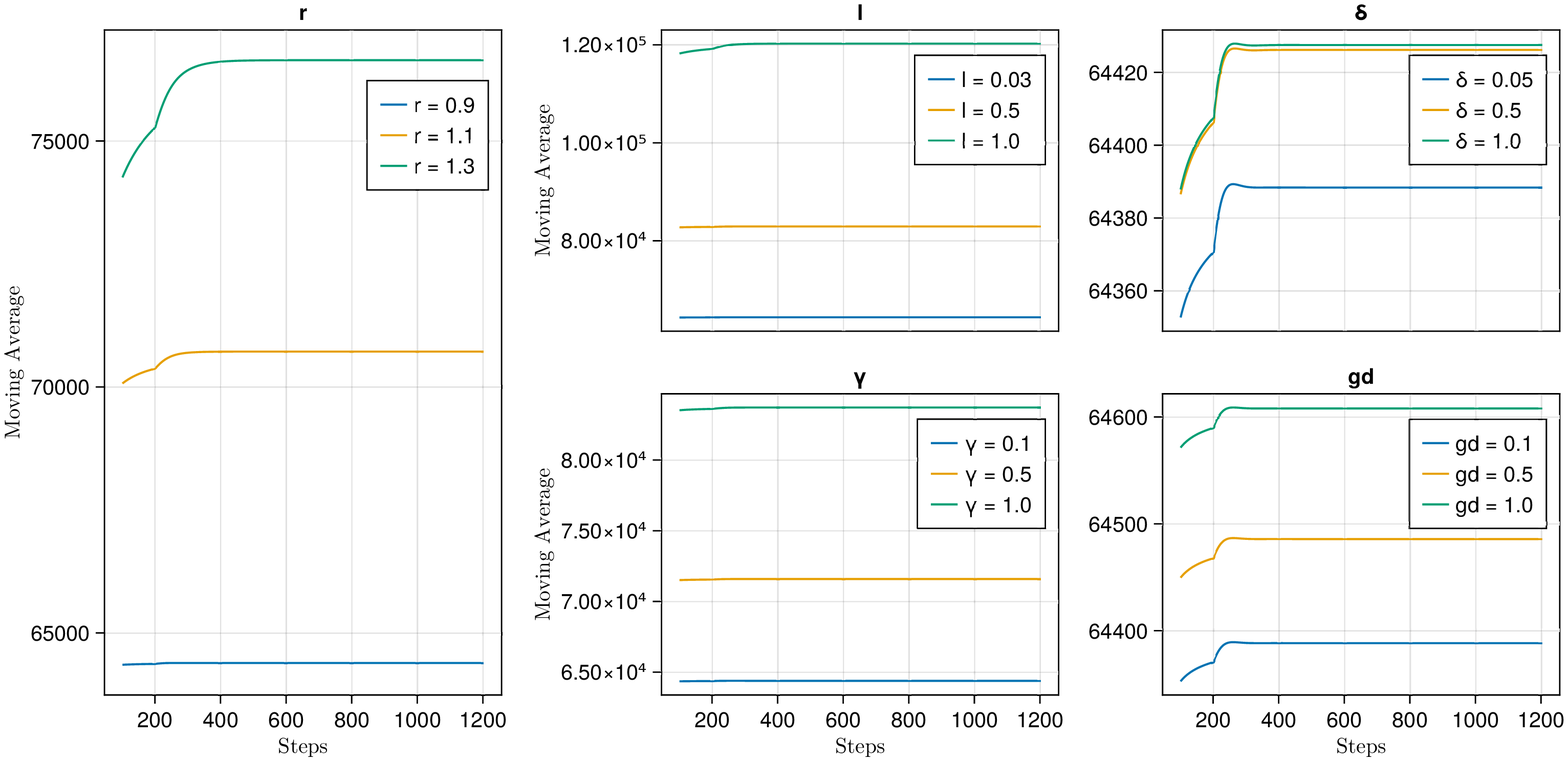}
    \caption{Sensitivity analysis per parameter - GDP dynamics. The baseline values used in the main simulations correspond to blue lines.}
    \label{fig:figureD1} %
\end{figure}

\begin{figure}[H]
    \centering
 \includegraphics[width = 0.7\paperwidth]{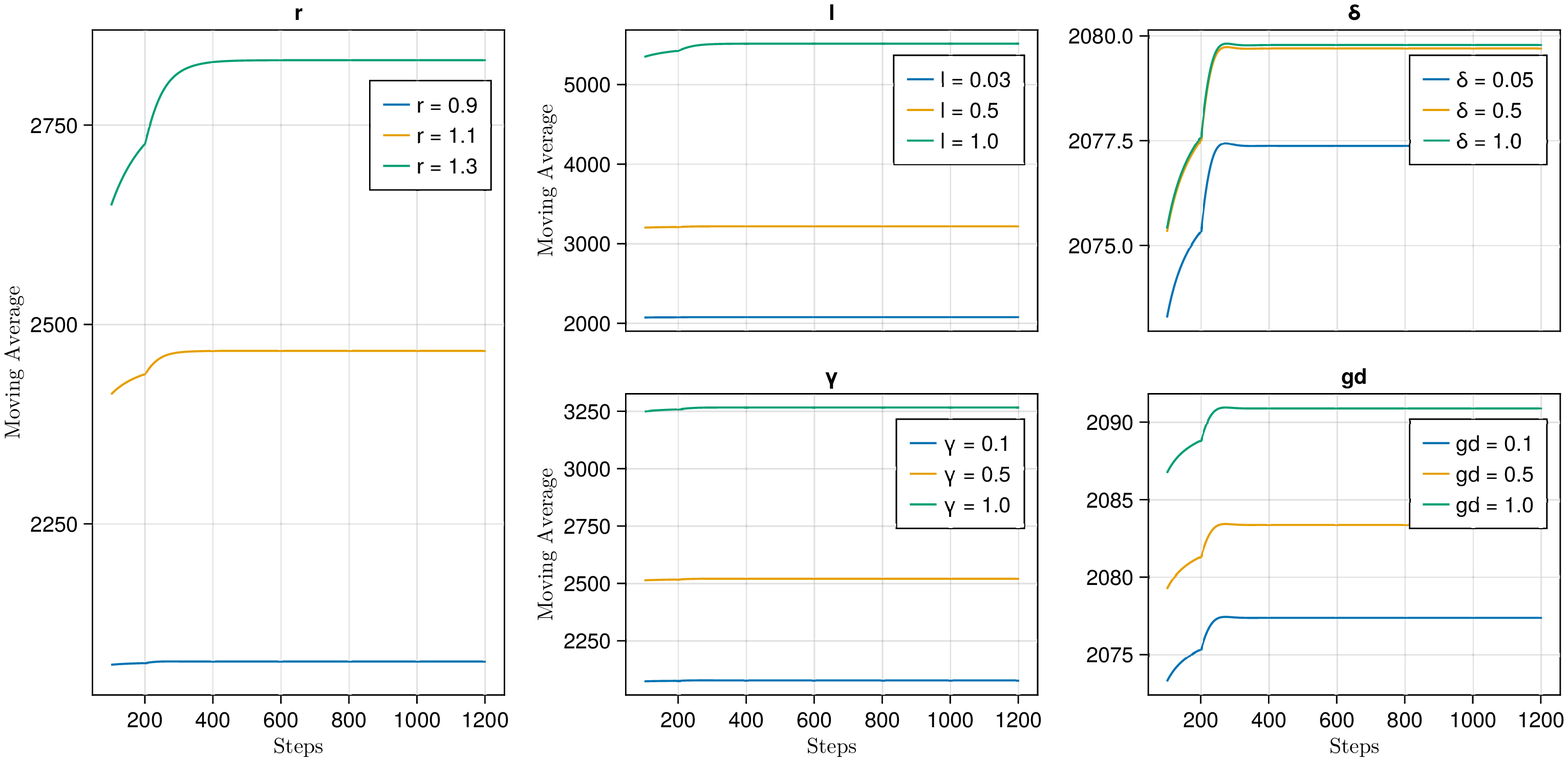}
    \caption{Sensitivity analysis per parameter - Overnight segment dynamics. The baseline values used in the main simulations correspond to blue lines.}
    \label{fig:figureD2} %
\end{figure}

\end{appendices}

\end{document}